\documentclass[aip,%
               jcp,%
               reprint,%
               citeautoscript,%
               amsfonts,%
               amssymb,%
               amsmath,%
               a4paper,%
               titlepage,%
               fleqn,%
               endfloats*%
               ]{revtex4-2}

\usepackage{graphicx}
\usepackage{color}
\usepackage{caption,subcaption}

\graphicspath{{./figs/}}

\begin{document}

%%%%%%%%%%%%%%%%%%%%%%%%%%%%%%%%%%%%%%%%%%%%%%%%%%%%%%%%%%%%%%%%%%%%%
%% If issues arise when submitting your manuscript, you may want to
%% un-comment the next line.  This provides information on the
%% version of every file you have used.
%%%%%%%%%%%%%%%%%%%%%%%%%%%%%%%%%%%%%%%%%%%%%%%%%%%%%%%%%%%%%%%%%%%%%

\newcommand*\mycommand[1]{\texttt{\emph{#1}}}
\newcommand{\hoch}[1]{$^{\text{#1}}$}
\newcommand{\tief}[1]{$_{\text{#1}}$}

% letters for footnote marks
\renewcommand*{\thefootnote}{\alph{footnote}}

% bold mathematical symbols
\def\bNul{{\boldsymbol{0}}}
\def\bsig{{\boldsymbol{\sigma}}}
\def\btau{{\boldsymbol{\tau}}}

% bold letters
\def\bfA{\boldsymbol{A}}
\def\bfB{\boldsymbol{B}}
\def\bfb{\boldsymbol{b}}
\def\bfC{\boldsymbol{C}}
\def\bfc{\boldsymbol{c}}
\def\bfD{\boldsymbol{D}}
\def\bfE{\boldsymbol{E}}
\def\bfF{\boldsymbol{F}}
\def\bff{\boldsymbol{f}}
\def\bfG{\boldsymbol{G}}
\def\bfg{\boldsymbol{g}}
\def\bfH{\boldsymbol{H}}
\def\bfI{\boldsymbol{I}}
\def\bfK{\boldsymbol{K}}
\def\bfL{\boldsymbol{L}}
\def\bfM{\boldsymbol{M}}
\def\bfP{\boldsymbol{P}}
\def\bfQ{\boldsymbol{Q}}
\def\bfR{\boldsymbol{R}}
\def\bfS{\boldsymbol{S}}
\def\bfT{\boldsymbol{T}}
\def\bfU{\boldsymbol{U}}
\def\bfV{\boldsymbol{V}}
\def\bfW{\boldsymbol{W}}
\def\bfX{\boldsymbol{X}}
\def\bfd{\boldsymbol{d}}
\def\bfs{\boldsymbol{s}}
\def\bft{\boldsymbol{t}}
\def\bfu{\boldsymbol{u}}
\def\bfv{\boldsymbol{v}}
\def\bfr{\boldsymbol{r}}
\def\bfx{\boldsymbol{x}}
\def\bNul{{\boldsymbol{0}}}
\def\bsig{{\boldsymbol{\sigma}}}
\def\bkap{{\boldsymbol{\kappa}}}
\def\bLam{\boldsymbol{\Lambda}}

% bra, ket, ...
\def\bra#1{\langle #1|}
\def\ket#1{|#1\rangle}
\def\ovlp<#1|#2>{\langle #1 | #2 \rangle}
\def\braket<#1|#2>{\langle #1 | #2 \rangle}
\def\komu<#1,#2>{\left[ #1 , #2 \right]}

% operators
\def\orb{\hat{\kappa}}
\def\cfg{\hat{S}}
\def\ham{\hat{H}}
\def\EOp{\hat{E}}
\def\E2Op{\hat{e}}
\def\htil{\hat{\widetilde{H}}}
\def\proj{\hat{P}}
\def\orb{\hat{\kappa}}
\def\cfg{\hat{S}}

% upper / lower case
\def\hoch#1{\textsuperscript{#1}}
\def\tief#1{\textsubscript{#1}}

% Greek letters
\def\gm{\mu}
\def\gn{\nu}
\def\gk{\kappa}
\def\gl{\lambda}
%

%% Each author should be given as a separate \author command.
%%
%% Corresponding authors should have an e-mail given after the author
%% name as an \email command. Phone and fax numbers can be given
%% using \phone and \fax, respectively; this information is optional.
%%
%% The affiliation of authors is given after the authors; each
%% \affiliation command applies to all preceding authors not already
%% assigned an affiliation.
%%
%% The affiliation takes an option argument for the short name.  This
%% will typically be something like "University of Somewhere".
%%
%% The \altaffiliation macro should be used for new address, etc.
%% On the other hand, \alsoaffiliation is used on a per author basis
%% when authors are associated with multiple institutions.
%%%%%%%%%%%%%%%%%%%%%%%%%%%%%%%%%%%%%%%%%%%%%%%%%%%%%%%%%%%%%%%%%%%%%

\author{Benjamin Helmich-Paris}

\email{helmichparis@kofo.mpg.de}
%\phone{+49 (0)208 3062164}
\affiliation{
 Max-Planck-Institut f{\"u}r Kohlenforschung,
 Kaiser-Wilhelm-Platz 1,
 D-45470 M{\"u}lheim an der Ruhr
}

%%%%%%%%%%%%%%%%%%%%%%%%%%%%%%%%%%%%%%%%%%%%%%%%%%%%%%%%%%%%%%%%%%%%%
%% The document title should be given as usual. Some journals require
%% a running title from the author: this should be supplied as an
%% optional argument to \title.
%%%%%%%%%%%%%%%%%%%%%%%%%%%%%%%%%%%%%%%%%%%%%%%%%%%%%%%%%%%%%%%%%%%%%
\title{
A trust-region augmented Hessian implementation for
 state-specific and state-averaged CASSCF wave functions
}

%%%%%%%%%%%%%%%%%%%%%%%%%%%%%%%%%%%%%%%%%%%%%%%%%%%%%%%%%%%%%%%%%%%%%
%% Some journals require a list of abbreviations or keywords to be
%% supplied. These should be set up here, and will be printed after
%% the title and author information, if needed.
%%%%%%%%%%%%%%%%%%%%%%%%%%%%%%%%%%%%%%%%%%%%%%%%%%%%%%%%%%%%%%%%%%%%%

\date{\today}


\begin{abstract}
In this work, we present a one-step second-order 
 converger for state-specific (SS) and state-averaged (SA)
 complete active space self-consistent field (CASSCF) wave functions.
Robust convergence is achieved through step restrictions
 using a trust-region augmented Hessian (TRAH) algorithm.
To avoid numerical instabilities,
 an exponential parametrization of variational configuration parameters is employed,
 which works with a nonredundant orthogonal complement basis.
This is a common approach for SS-CASSCF and is extended to SA-CASSCF wave functions, in this work.
Our implementation is integral direct and based on intermediates that are
 formulated either in the sparse atomic-orbital or small active molecular-orbital basis.
Thus, it benefits from a combination with efficient integral decomposition techniques,
 such as the resolution-of-the-identity or the chain-of-spheres for exchange approximations.
This facilitates calculations on large molecules such as a Ni(II) complex with 231~atoms
 and 5154~basis functions.
The runtime performance of TRAH-CASSCF is competitive
 with other state-of-the-art implementations of approximate and full second-order algorithms.
In comparison with a sophisticated first-order converger,
 TRAH-CASSCF calculations usually take more iterations to reach convergence and, thus, have longer runtimes.
However, TRAH-CASSCF calculations still converge reliably to a true minimum even if the first-order algorithm fails.
\end{abstract}

\maketitle

%--------------------------------------------------------------------%
\section{Introduction} \label{sec:intro}
%--------------------------------------------------------------------%

% (1) MR methods when needed
Multi-reference (MR) methods become indispensable if the leading 
  wave function expansion coefficients are similar in magnitude.
Though this situation rarely occurs when studying most closed-shell molecules at their equilibrium geometry,
 it is of high relevance in many applications that involve
(i)  potential curves or surfaces with stretched or broken covalent bonds,
(ii) low-lying electronically excited states that are reached by double-electron excitations from the ground state --- as it occurs in polyenes,
(iii) conical intersections and avoided crossings of potential surfaces,
 and (iv) --- probably most import --- transition-metal complexes (TMC) with
 nearly degenerate open-shell ground states or with magnetically coupled centers.

% (2) CASSCF
The complete active space self-consistent field (CASSCF) method is
 the most frequently used MR method. 
The basic idea is fairly simple.
First, a set of chemically relevant active orbitals and electrons needs to be defined
 before the calculation which is referred to as active space (AS). 
Then, the CASSCF wave function is expanded in all combinatorially possible spin-adapted 
 determinants --- also known as configuration state functions (CSF) ---
 that are obtained by distributing the AS electrons among the AS orbitals.
With such a configuration-interaction (CI) expansion, the Schr{\"o}dinger equation is solved by
 minimizing the energy variationally to find optimal molecular orbital (MO) and CI coefficients.
The CASSCF solution is often required for more accurate --- though much more costly --- MR wave function methods 
 that account for both static and dynamic electron correlation.\cite{Andersson1990,*Andersson1992,*Finley1998,Angeli2001,*Angeli2001b,*Angeli2002,Angeli2004,Werner1988,*Shamasundar2011,Hanauer2011,*Hanauer2012,*Aoto2016}
   
% (3) numerical challenges
Finding the proper minimum solution imposes a numerical challenge for many CASSCF calculations.
Those issues are mainly caused by the division of the orbital space
 into inactive orbitals with either double- or zero-electron occupation
 and active orbitals for which the CI expansion is made.
In the course of the optimization process, orbitals may perpetually change their roles
 often leading to convergence difficulties and many possible minimum solutions.
In particular, this is observed if strongly and/or weakly occupied orbitals with occupation numbers 
 near 2.0 and 0.0, respectively, are part of the active space.
  
% (4) remedy
These issues can be avoided to some extent by appropriate initial MOs 
 that are coherent with the choice of active space.
Automated procedures like the \textit{atomic valence active space}\cite{Sayfutyarova2017} (AVAS)
 or the \textit{$\mathit{\pi}$ orbital space}\cite{Sayfutyarova2019} (PiOS)
 can provide this and are extensively used in the present work.
Another approach to avoid CASSCF convergence issues is employing
 robust second-order convergers that incorporate information from the gradient and the Hessian  
 into the optimization procedure.
Since this subject was recognized in the early days of multiconfigurational (MC) SCF theory,
 there is currently a long record of second-order (and beyond) MCSCF implementations,\cite{Yaffe1976,Yeager1979,*Dalgaard1979,*Olsen1983,Siegbahn1980,*Siegbahn1981,Lengsfield1980,*Lengsfield1981,*Lengsfield1982,Werner1980,*Werner1981,*Werner1985,Igawa1982,Jensen1984,*Jensen1986,*Jensen1987,Jensen1996,Hedegaard2018,*Lipparini2016,*Reynolds2018,Sun2017,Ghosh2008,*Ma2017,Kreplin2019,Kreplin2020,Nottoli2021}
 each with its merits and shortcomings.
%An thorough comparison is beyond the scope of the present work and we focus on putting
% our own work in the context of previous works.

% TRAH-CASSCF
In the present work, we present the trust-region augmented Hessian (TRAH) method
 for state-specific (SS) and state-averaged (SA) CASSCF wave functions,
 which builds upon our previous work on single-reference TRAH-SCF.\cite{Helmich-Paris2021}
As Yeager, J{\o}rgensen, and Dalgaard\cite{Yeager1979,*Dalgaard1979,*Olsen1983},
 we employ a full second-order CASSCF expansion of the energy by means of all variational parameters.
The variational parameters for the MO and CI coefficients are determined in one step.
Additionally, the TRAH algorithm introduces a step restriction to avoid
 divergence when far from the minimum solution.
Such a step restriction was already introduced for the norm-extended
 optimization (NEO) of Jensen and J{\o}rgensen.\cite{Jensen1984,*Jensen1986,*Jensen1987,Nottoli2021}
In contrast to previous NEO implementations, 
 the unmodified configuration-configuration Hessian enters the augmented Hessian
 or Newton-Raphson (NR) equations. 
This block of the Hessian is singular when employing a linear parametrization for
 the CI coefficient updates, which are expanded in the entire CSF basis.
Instead, we use an exponential parametrization with state-rotation parameters expanded
 in a nonredundant orthogonal complement basis, as customary 
 for the full second-order NR implementations\cite{Yeager1979,*Dalgaard1979,*Olsen1983}
 and MCSCF response theory implementations of numerous properties.\cite{Olsen1985,*Helgaker1986,*Joergensen1988,*Hettema1992}
Additionally, we show in this work how to extend this formalism
 to the second-order optimization of SA-CASSCF wave functions, which are, in practice,
 as important as SS calculations.
This is elaborated on in Sec.\ \ref{sec:theory}.

Our integral-direct atomic orbital-based (AO) implementation of TRAH-CASSCF
 is presented in Sec.\ \ref{sec:implement}.
There, we focus on an efficient computation of AO-Fock matrices and integrals
 with state-of-the-art integral decomposition techniques
 and also discuss how to reduce the number of iterations by choosing
 beneficial orbital representations and by sophisticated preconditioners.

The robustness and efficiency of our TRAH-CASSCF implementation is eventually
 verified in Sec.\ \ref{sec:results} by investigating the convergence and the runtime in prototypical
 applications that involve
 aromatic molecules,
 open-shell TMC with (quasi-)degenerate ground states,
 and magnetically coupled systems.

%--------------------------------------------------------------------%
\section{Theory} \label{sec:theory}
%--------------------------------------------------------------------%

%-----------------------------------------------------------------------%
\subsection{CASSCF wave function ansatz}
%-----------------------------------------------------------------------%

The CASSCF wave function $\ket{0}$ is expanded in all possible configuration state functions (CSF)
$\ket{\Phi_I}$ 
\begin{align} \label{eq:cas-wf}
 \ket{0} &= \sum_I C_I \ket{\Phi_I}
\end{align}
that are obtained by distributing a given number of active electrons among a given
number of active MOs that are usually of valence type.
The CASSCF solutions can then be obtained by inserting Eq.\ \eqref{eq:cas-wf} into
 the nonrelativistic Schr{\"o}dinger equation
\begin{align} \label{eq:ss-cas-ener}
 E = \min_{\bkap,\bfS} \bra{0} \ham \ket{0}
\end{align}
 and by minimizing $E$ with respect to
 variations in the MO $\bkap$
 and CI $\bfS$ coefficients.
The unit norm of the wave function is assumed here
 and in the following.
The nonrelativistic Hamiltonian in Eq.\ \eqref{eq:ss-cas-ener} is given
 in its second-quantized from by
\begin{align}
 \ham &= \sum_{pq} h_{pq} \EOp_{pq} + \frac{1}{2} \sum_{pqrs} (pq|rs)  \E2Op_{pqrs} \\
 \E2Op_{pqrs} &=  \EOp_{pq} \EOp_{rs} - \delta_{qr}  \EOp_{ps}
\end{align}
 with $p$, $q$, $r$, and $s$ denoting general MOs,
 $h_{pq}$ and $(pq|rs)$ are the usual real, one- and two-electron integrals, respectively.
The final CASSCF wave function parameters, i.e.\ the MO and CAS-CI coefficients $C_I$,
 are obtained from the minimum energy in Eq.\ \eqref{eq:ss-cas-ener}.
An average of multiple states $i$ with a fixed weighting factor $w_i$ can be found
 in the same way,
\begin{align} \label{eq:sa-cas-ener}
 E = \min_{\bkap,\bfS} \sum_i w_i  \bra{0_i} \ham \ket{0_i} \text{,}
\end{align}
 which defines the SA-CASSCF wave function model.
For a multitude of states in Eq.\ \eqref{eq:sa-cas-ener},
 the same orbitals are used.

Variations in the MO coefficients are conveniently described by
means of real anti-symmetric orbital rotation parameters $\bkap$
\begin{align}
 \orb        &= \sum_{p>q} \kappa_{pq} \EOp_{pq}^- \\
  \kappa_{pq} &= - \kappa_{qp} \\
 \EOp_{pq}^- &= \EOp_{pq} - \EOp_{qp} \text{.}
\end{align}
The anti-symmetric orbital rotations enter a unitary operator 
\begin{align}
 \widetilde{\ket{0}} & = \exp(-\orb)\ket{0} \label{eq:exporb}
\end{align}
that connects the orbital
part of the initial $\ket{0}$ with the optimized $\widetilde{\ket{0}}$ SCF-type wave function.
The exponential parametrization in Eq.\ \ref{eq:exporb}
 leads to a simple evaluation by means of second-quantized operators
 and ensures orthonormality of the MOs provided that the initial MO
 set is orthonormal as well.

For the optimization of the CI coefficients, there are two common parametrizations.
Most common is the linear parametrization\cite{Jensen1994}
\begin{align}
 \widetilde{\ket{0}} & = \frac{  \ket{0} + \proj \ket{S} } { \sqrt{ 1 + \bra{S} \proj  \ket{S} } } \text{,} \label{eq:ci-lin-para}
\end{align}
 which works with an expansion in the CSF basis $\{ \Phi_I \}$ (like Eq.\ \eqref{eq:cas-wf}).
 However, the CSF basis is redundant in this context because
 it spans both the current solution $\ket{0}$ and variations in the CI parameter space $\ket{S}$ 
 which should be mutually orthogonal. 
 Thus, a projector is required
\begin{align}
 \proj &= 1 - \sum_j \ket{0_j}\bra{0_j}
\end{align}
 that removes $\ket{0}$ out of $\ket{S}$ explicitly and ensures that $\braket< 0 | S > = 0$.

\begin{figure}
 \centering
  \scalebox{0.50}{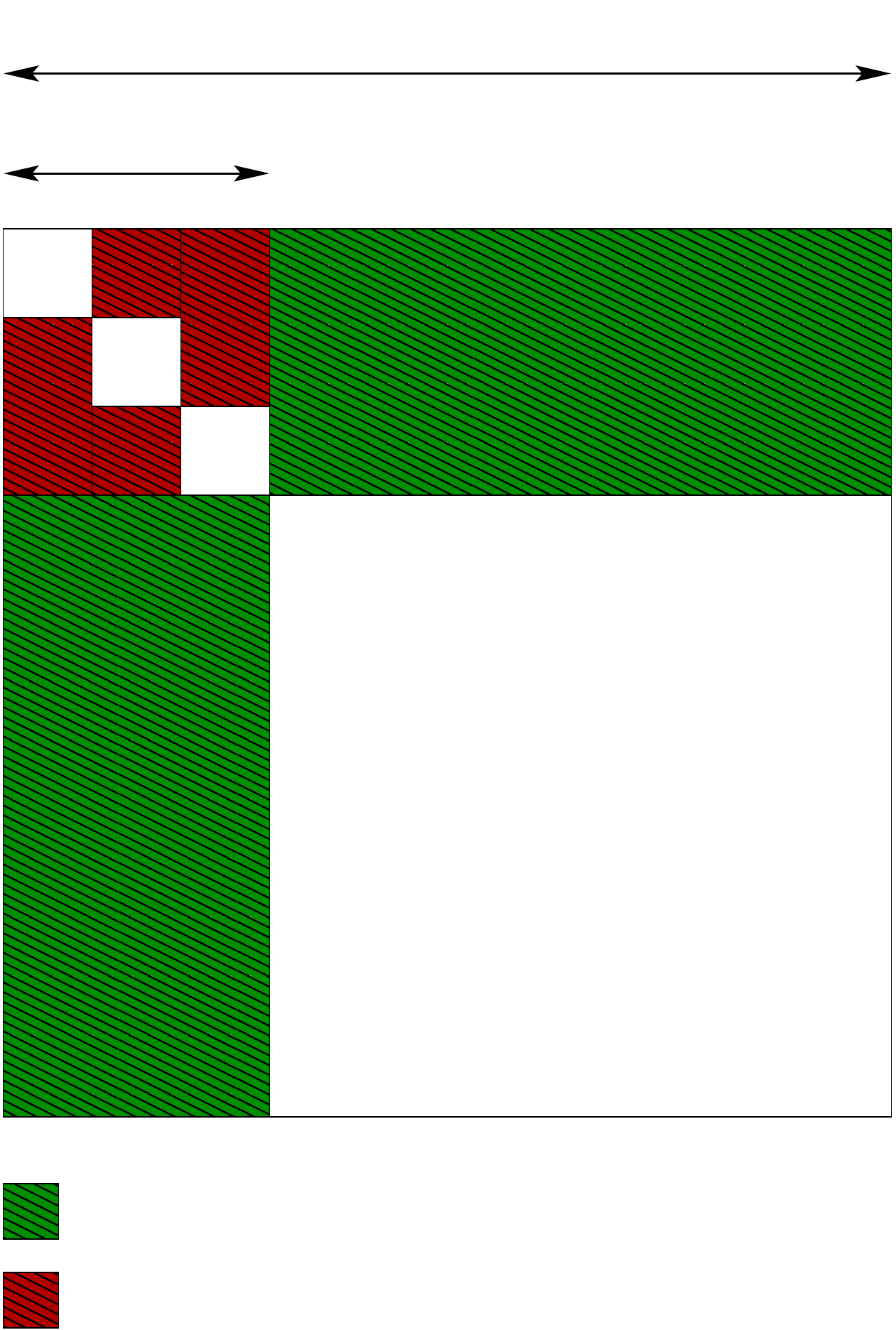}
 \caption{Schematic structure of the anti-symmetric state-rotation parameters $\bfS$.}
 \label{fig:1}
\end{figure}

The exponential parametrization\cite{Yeager1979,*Dalgaard1979,*Olsen1983,Olsen1985}
 \begin{align}
  \widetilde{\ket{0_j}} &= \exp( -\hat{S}  ) \ket{0_j} \label{eq:exp-s-op} \\
  \hat{S} &= \sum_{k>j} S_{kj} \left( \ket{k}\bra{0_j} - \ket{0_j}\bra{k} \right) \label{eq:s-op}
 \end{align}
 works directly in a nonredundant orthogonal complement basis $\{ \ket{k} \}$
 that has by construction no overlap with the current CI solutions $\{ \ket{0_j} \}$.
As for $\bkap$ the state-rotation parameters $\bfS$ are anti-symmetric
 \begin{align}
  S_{kj} = - S_{jk}
 \end{align}
 with a structure illustrated in Fig.\ \ref{fig:1}.
Note that in Eq. \eqref{eq:s-op} $k$ runs over all states included in the average $\ket{0_k}$ with $k>j$
 as well as the orthogonal complement space $\{ \ket{K} \}$, which if unified with the SA 
 states $\{ \ket{0_j} \}$ gives again the CSF space,
 i.e.\ $ \{ \ket{0_k} \} \cup \{ \ket{K} \}= \{ \Phi_K\} $.
It is shown in Sec.\ \ref{sec:ci-bas}  how this nonredundant basis is 
 obtained and employed for SS- and SA-CASSCF
 calculations.

%-----------------------------------------------------------------------%
\subsection{TRAH Optimization of CASSCF wave functions}
%-----------------------------------------------------------------------%

The restricted-step second-order energy optimization can be applied whenever
 the minimum energy $E$ can be expanded in a Taylor series 
\begin{align}
 E &= E_0 + {\bfx}^t \bfg + \frac{1}{2} {\bfx}^t \bfH {\bfx} + \ldots \label{eq:e-taylor-o2}
\end{align}
around a current expansion point (CEP) the represents the current wave functions parameters.
The electronic energy $E_0$,
 and its first (gradient $\bfg$) and second derivatives (Hessian $\bfH$) with respect to all variations $\bfx$
 in the wave function parameters must then be computed.
In the TRAH algorithm,\cite{Hoeyvik2012a,Helmich-Paris2021,Nottoli2021b} 
  the update of the current parameters is obtained by searching for the 
  lowest eigenvalue and its corresponding eigenvector of the scaled
  augmented Hessian (AH) matrix
\begin{align}
  \begin{pmatrix}
   0 & \alpha \, {\bfg}^t \\
\alpha \, {\bfg} &  \bfH 
  \end{pmatrix}
  \begin{pmatrix}
 1 \\
 \bfx(\alpha)
  \end{pmatrix}
=
 \mu \,
  \begin{pmatrix}
 1 \\
 \bfx(\alpha)
  \end{pmatrix}
\label{eq:trah-eig}
 \end{align}
 by using an iterative Davidson-type algorithm.\cite{Davidson1975}
The scaling parameter $\alpha$ is determined together with $\mu$ and $\bfx(\alpha)$ in the 
 iterative diagonalization of \eqref{eq:trah-eig} known as the micro-iterations
 and is determined by enforcing the update vector
\begin{align}
 \bfx &= \frac{1}{\alpha} \bfx(\alpha)
\end{align}
to lie within a trust sphere  with radius $h$
\begin{align}
 ||\bfx ||^2 \le h^2 \text{.}
\end{align}
The trust radius $h$ is adjusted dynamically after the wave function parameter update has been
 made by comparing the actual energy change with the predicted second-order energy estimate,
 which is known as Fletcher's algorithm.\cite{Fletcher1987}
Since the expansion in Eq.\ \eqref{eq:e-taylor-o2} is truncated after second-order terms,
 updating the CEP by $\bfx$ will only move the CEP directly into the minimum if the CEP is already
 close to the minimum.
Therefore, the eigenvalue equations \eqref{eq:trah-eig} and the parameter update must be performed
 multiple times to reach convergence what defines the so-called macro-iterations.
More algorithmic details can be found in our recent work on the TRAH-SCF implementation.\cite{Helmich-Paris2019}

%-----------------------------------------------------------------------%
\subsection{CASSCF energy, gradient, and Hessian}\label{sec:egh}
%-----------------------------------------------------------------------%
Let us start with discussing the configuration gradient and
 Hessian in a the linear, redundant CI parametrization (Eq.\ \eqref{eq:ci-lin-para}) to motivate why we prefer the exponential,
 nonredundant parametrization.
For the sake of simplicity, we discuss in this context only the state-specific case.
First, a CASSCF energy Lagrangian is introduced that accounts for the normalization
 of the optimized wave function $\widetilde{\ket{0}}$ and ignores orbital rotations
 for reasons of notational convenience
\begin{align}
 L(\bfc, E) &=  \left( \bra{S} \proj + \bra{0} \right) \, \ham \, \left( \ket{0}  + \proj \ket{S} \right)
- E \bra{S} \proj \ket{S} \text{.}
\end{align}
Since we expand update vectors $\bfx$ and their respective trial vectors in the CSF basis,
 we have to ensure that, whenever the current solution $\bfC$ is parallel to $\bfx$,
 (i) the gradient is still orthogonal to $\bfC$ and (ii) the Hessian is nonsingular
 when being multiplied by $\bfC$.

The configuration gradient in the linear parametrization (Eq.\ \eqref{eq:ci-lin-para}) reads
\begin{align}
 {\bfg}^c =   \left. \frac{ \partial L }{ \partial S_{I} } \right|_*
 &=  \bra{I} \proj \ham \ket{0} + \bra{0} \ham \proj \ket{I} \\
 &=  2 \left( \bra{I} \ham \ket{0} - E \, C_I \right)
\end{align}
with the current energy given by
\begin{align}
 E &= \bra{0} \ham \ket{0} \text{.}
\end{align}
The gradient is orthogonal to the current solution $\bfx = \{ \bNul, \bfC \}$ 
\begin{align}
 ({\bfg})^T \, \bfx = ({\bfg^c})^T \, \bfC = 2 \left( \bra{0} \ham \ket{0} - E  \right) = 0 \label{eq:g-dot-x}
\end{align}
with a given set of MOs.
Due to the requirement that the current solution and its update vector must be orthogonal, 
 the linear, redundant parametrization in Eq.\ \eqref{eq:ci-lin-para}  is eligible
 for a first-order algorithm, in which the variations in the CI space $\bfS$ are solely determined
 by the gradient.

Using again the linear, redundant parametrization, the CI Hessian reads
\begin{align}
  \left. \frac{ \partial^2 L }{ \partial S_I \partial S_J } \right|_*
 &= 2 \left( \bra{I} \, \proj \, H \, \proj \, \ket{J}  - E \, \bra{I} \, \proj \, \ket{J} \right) \\
 &= 2 \left( \bra{I} H \ket{J}  - E \, \delta_{IJ} \right) 
 - C_I \, g^c_J - g^c_I \, C_J \label{eq:ci-red-hess}
\end{align}
When computing a CI sigma vector of \eqref{eq:ci-red-hess} with $\bfC$ 
\begin{align}
  {\bfH}^{cc} \, {\bfC} &= 2 \left( \bra{I} H \ket{C}  - E \, C_I \right) - {\bfg}^c = \bNul
\end{align}
 and making use of Eq.\ \eqref{eq:g-dot-x}, a null vector is obtained.
Without taking further actions, the linear, redundant CI parametrization will inevitably lead to numerical instabilities
 whenever solving linear or eigenvalue equations with the Hessian in \eqref{eq:ci-red-hess}.
Note that in the NEO second-order optimization algorithm\cite{Jensen1984,Jensen1986}
 a modified CI Hessian occurs that is not singular when working with the linear, 
 redundant parametrization.
Thus, these numerical instabilities do not occur with NEO.

In the current work, we will work instead with the exponential, nonredundant parametrization
 of the configuration update Eqs.\ \eqref{eq:exp-s-op}.
For the latter, the state-specific minimum energy
\begin{align}
 E^i &= \widetilde{ \bra{0_i} } \ham \widetilde{ \ket{ 0_i } } \\
    &=  \bra{0_i} \exp(\cfg) \exp( \orb) \ham \exp(-\orb) \exp(-\cfg) \ket{ 0_i } 
\end{align}
 can be easily evaluated by using the Baker-Campbell-Hausdorff expansion
 of the double exponential parametrization up to second order
\begin{align}
 E^i \approx \bra{0_i} &\ham + \komu< \orb, \ham > + \komu< \cfg, \ham > \notag \\
& \frac{1}{2} \komu< \orb, \komu< \orb, \ham > >
+             \komu< \cfg, \komu< \orb, \ham > >
+ \frac{1}{2} \komu< \cfg, \komu< \cfg, \ham > >
 \ket{ 0_i }  \label{eq:ener-exp}
\end{align}
Note that we imply that operators for orbital $\orb$ and configuration rotations  $\cfg$
 commute which requires certain restrictions on the allowed rotations
 between active orbitals.
In case of CASSCF, these redundancies between the orbital and configuration rotation part
 are handled with ease by omitting all rotations among active orbitals.
%\begin{align}
% E = \sum_i w_i \bra{0_i} \ham \ket{0_i}
%\end{align}
% does not conis 

The SA-CASSCF energy gradient terms are easily obtained from Eq.\ \eqref{eq:ener-exp}
 and are given by
\begin{align}
 {\bfg}^o &= \left. \frac{ \partial E }{ \partial \kappa_{pq} } \right|_*
 = \sum_i w_i \bra{0_i} \left[ \EOp^-_{pq}, \ham \right] \ket{0_i} \label{eq:g-orb} \\
 {\bfg}^{c_e} &=   \left. \frac{ \partial E }{ \partial S_{Kj} } \right|_*
 =  -2 w_j \bra{K} \ham \ket{0_j} \label{eq:g-cfg-e}\\
 {\bfg}^{c_i} &=   \left. \frac{ \partial E }{ \partial S_{kj} } \right|_*
 =  2 ( w_k - w_j ) \, \bra{0_k} \ham \ket{0_j} \label{eq:g-cfg-i} \text{.}
\end{align}
When taking derivatives, we distinguish between (internal) rotations within SA states $0_j$ and $0_k$ (Eq.\ \eqref{eq:g-cfg-i})
 and (external) rotations between an SA state $0_j$ and the orthogonal complement states $K$ (Eq.\ \eqref{eq:g-cfg-e}).
As becomes obvious from Eq.\ \eqref{eq:g-cfg-i},
 there are a few scenarios at which the internal state rotations vanish:
 (i) all state weights $w_i$ are equal,
 (ii) there is only a single state for each spin or point-group symmetry \textit{irrep}, and
 (iii) the CI coefficients are transformed in a basis that keeps the state-interaction matrix
\begin{align}
\bra{0_i} \ham \ket{0_j} = \delta_{ij} E_j \label{eq:inter-state}
\end{align}
 diagonal.

As for the gradients, the SA-CASSCF energy Hessian is easily deduced from Eq.\ \eqref{eq:ener-exp}
 and reads
\begin{align}
 {\bfH}^{oo} &=  \left. \frac{ \partial^2 E }{ \partial \kappa_{pq} \, \partial \kappa_{rs} } \right|_* \notag \\
  &=  \frac{1}{2} \left( 1 + \hat{\mathcal{P}}_{pq,rs} \right) \sum_i w_i \, \bra{0_i} \left[ \EOp^-_{pq}, \left[ \EOp^-_{rs}, \ham \right] \right] \ket{0_i} \label{eq:h-orb-orb} \\[0.5em]
 {\bfH}^{c_eo} &=   \left. \frac{ \partial^2 E }{ \partial S_{Kj} \, \partial \kappa_{pq} } \right|_* \notag \\
 &=  - w_j \left( \bra{K} \left[ \EOp^-_{pq}, \ham \right] \ket{0_j} + \bra{0_j} \left[ \EOp^-_{pq}, \ham \right] \ket{K} \right) \label{eq:h-cfg-orb} \\[0.5em]
 {\bfH}^{c_ec_e} &=   \left. \frac{ \partial^2 E }{ \partial S_{Kj} \, \partial S_{Mn} } \right|_* \notag \\
 &=  2 w_j \left( \delta_{jn} \bra{K} \ham \ket{M} -  \delta_{KM} \bra{0_j} \ham \ket{0_n} \right) \label{eq:h-cfg-cfg}
\end{align}
\begin{align}
 {\bfH}^{c_io} &=   \left. \frac{ \partial^2 E }{ \partial S_{kj} \, \partial \kappa_{pq} } \right|_*  \notag \\
 &=  -(w_j- w_k) \, \left( \bra{0_k} \left[ \EOp^-_{pq}, \ham \right] \ket{0_j} + \bra{0_j} \left[ \EOp^-_{pq}, \ham \right] \ket{0_k} \right) \label{eq:h-cfg-i-orb} \\[0.5em]
 {\bfH}^{c_ic_e} &=   \left. \frac{ \partial^2 E }{ \partial S_{kj} \, \partial S_{Mn} } \right|_*  \notag \\
 &=  2 (w_j-w_k) \left( \delta_{jn} \bra{0_k} \ham \ket{M} + \delta_{kn} \bra{0_j} \ham \ket{M}
\right) \label{eq:h-cfg-i-cfg-e}\\[0.5em]
 {\bfH}^{c_ic_i} &=   \left. \frac{ \partial^2 E }{ \partial S_{kj} \, \partial S_{mn} } \right|_* \notag \\
 &=  2 (w_j-w_k) \left( \delta_{jn} \bra{0_k} \ham \ket{0_m} -  \delta_{km} \bra{0_j} \ham \ket{0_n} \right. \notag \\
 &+ \phantom{2 (w_j-w_k)}\left. \delta_{kn} \bra{0_j} \ham \ket{0_m} -  \delta_{jm} \bra{0_k} \ham \ket{0_n} \right)\label{eq:h-cfg-i-cfg-i}
\end{align}
Some of the internal-state rotation blocks of the Hessian above scale quadratically
 with the number of SA states.
Concerning the computational costs of these terms, this is surely unproblematic
 for the state-interaction matrices (Eq.\ \eqref{eq:inter-state}) but becomes
 at some point time-determining for the mixed internal state - orbital rotation derivative \eqref{eq:h-cfg-i-orb}.
The latter necessitates the computation of $N_{\text{SA}}^2$
 active Fock matrices\cite{Snyder2017} in every macro-iteration which is one 
 the most time-consuming steps for CASSCF methods (\textit{vide infra}).
Thus, it is more reasonable to omit all inter state-rotation blocks of the gradient (Eq.\ \eqref{eq:g-cfg-i}) and
 Hessian (Eqs.\ \eqref{eq:h-cfg-i-orb} -- \eqref{eq:h-cfg-i-cfg-i}) by enforcing them to be redundant.
As discussed for the gradient, this is automatically the case for equal-weight calculations,
 which are the most practically relevant SA calculations.
If the state weights differ and if there is more than a single state per spin or PG symmetry \textit{irrep},
 we transform at the CEP the CI coefficients in a basis that diagonalizes the SI Hamiltonian (Eq.\ \eqref{eq:inter-state}).
%

%--------------------------------------------------------------------%
 \subsection{An integral-direct implementation}
%--------------------------------------------------------------------%

%--------------------------------------------------------------------%
\subsubsection{Gradient}
%--------------------------------------------------------------------%

\begin{figure}
 \centering
 \includegraphics[height=0.50\textheight]{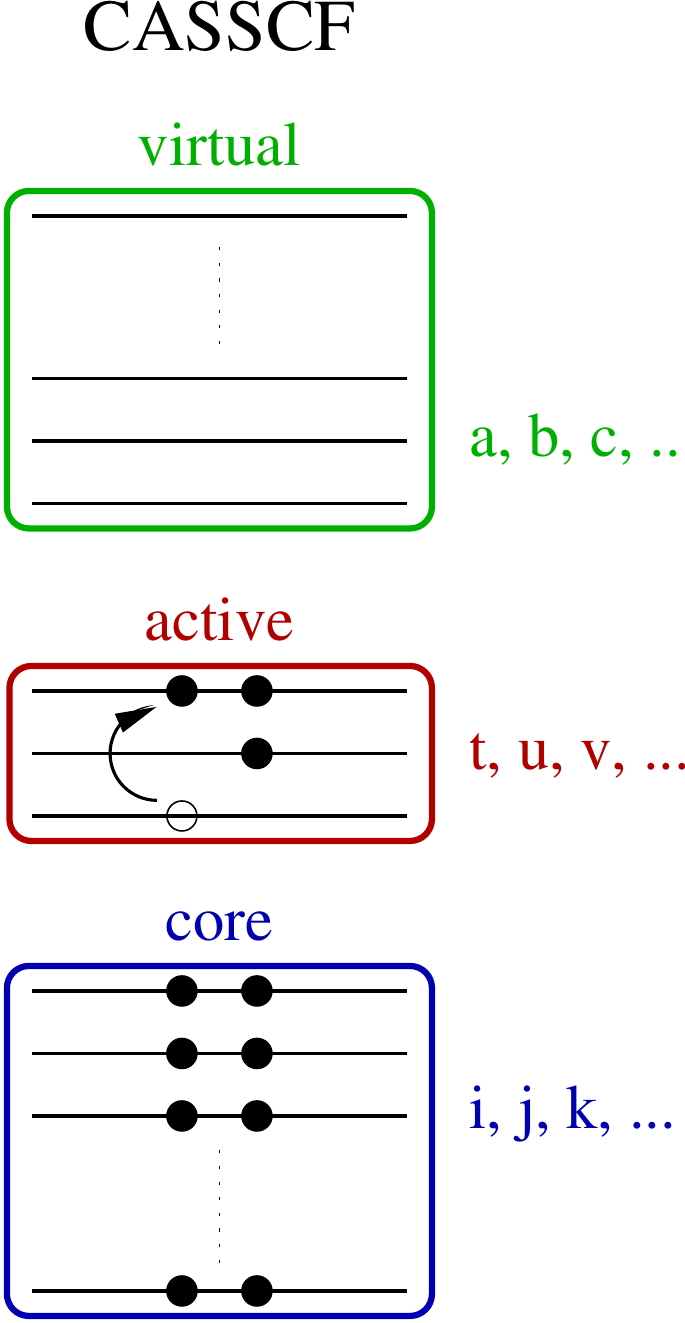}
 \caption{Schematic structure of MO energy level diagram with default index labels 
 for core (inactive), active, and virtual MO subspaces.}
 \label{fig:2}
\end{figure}

In CASSCF the only nonredundant orbital-rotation parameters involve 
 virtual-inactive, active-inactive, and virtual - active orbital pairs,
\begin{align}
 \orb &= 
 \sum_{ai} \kappa_{ai} \EOp^-_{ai}
+\sum_{ti} \kappa_{ti} \EOp^-_{ti}
+\sum_{at} \kappa_{at} \EOp^-_{at}
\text{.} \label{eq:k-non-red}
\end{align}
The naming convention for MO subspace labels is introduced in Fig.\ \ref{fig:2}.
It would be sufficient to compute only for those three blocks the orbital gradient
 and the orbital sigma vectors.
However, the active-active orbital gradient block occurs as an intermediate in the
  sigma vector computation.\cite{Jensen1986}
All four nonredundant orbital gradient blocks are then given by
\begin{align}
 g^o_{ai} &= -2 \left( 2 F^I_{ia} \braket<0|0> + 2 F^A_{ia} \right) \label{eq:g-ai}  \\[.4em]
 g^o_{ti} &= -2 \left( 2 F^I_{it} \braket<0|0> + 2 F^A_{it} - \sum_v D_{tv} F_{iv}^I - Q_{it} \right) \label{eq:g-ti} \\[.4em]
 g^o_{at} &= -2 \left( \sum_v F^I_{va} D_{tv} + Q_{at} \right) \label{eq:g-at} \\[.4em]
 g^o_{ut} &= -2 \left( \sum_v F^I_{vu} D_{vt} +  Q_{ut} - \sum_v D_{uv} F_{tv}^I - Q_{tu} \right)\label{eq:g-ut}
\end{align}
where the intermediates from Tab.\ \ref{tab:1} are employed.
In case of SA calculations, the weighted state sum is incorporated into the 
 overlap and one- and two-particle density matrices $D_{tu}$ and $d_{tuvw}$, respectively.

The configuration gradient in the nonredundant basis Eq.\ \eqref{eq:g-cfg-e}
is computed as
\begin{align}
 &g_{Ii} = -2 \, w_i \sum_{I'} \mathcal{U}_{II'} \, \bra{\Phi_{I'}} \ham \ket{ 0_i} \label{eq:g-cfg}
\end{align}
and requires a CI sigma vector in the CSF basis  with the current CI solution
\begin{align}
  &\bra{\Phi_I} \ham \ket{ 0_i} = \sum_J \bra{\Phi_I} \ham \ket{ \Phi_J }  C^i_J \notag \\
 &\bra{\Phi_I} \ham \ket{ \Phi_J } = E^c \delta_{IJ} + \sum_{tu} h'_{tu}A^{tu}_{IJ} \notag \\
 &+ \frac{1}{2} \sum_{tuvw} (tu|vw) \sum_K A^{tu}_{IK} A^{vw}_{KJ}  \label{eq:ci-sigma}
\end{align}
with subsequent 
CI basis transformation by means of $\boldsymbol{\mathcal{U}}$ (\textit{vide infra}).
The intermediates in Eq.\ \eqref{eq:ci-sigma} are compiled in Tab.\ \ref{tab:1},
\begin{align}
A^{tu}_{IJ} = \bra{ \Phi_I } \EOp_{tu} \ket{ \Phi_J }
\end{align}
 are the coupling coefficients\cite{Siegbahn1984} that can be computed
 by various methods.\cite{Knowles1984,Manne1985,Golebiewski1985,Shavitt1977,*Shavitt1978,Olsen1988b}

%--------------------------------------------------------------------%
\subsubsection{Sigma vectors}
%--------------------------------------------------------------------%

Products of the Hessian matrix (Eqs. \eqref{eq:h-orb-orb}, \eqref{eq:h-cfg-orb}, and \eqref{eq:h-cfg-cfg}) with
 trial vectors are commonly referred to as sigma vectors. 

The orbital-orbital sigma vector can be easily formulated 
 in terms of orbital gradients 
\begin{align}
 & \sigma_{pq}( \bkap) = {\bfH}^{oo}\, {\bkap} \\
 &= \sum_j w_j \left( \bra{0_j} \left[ \EOp^-_{pq}, \htil \right] \ket{0_j} - \frac{1}{2} \bra{0_j} \left[ \left[ \EOp^-_{pq}, \hat{\kappa} \right], \ham \right] \ket{0_j} \right) \\[0.5em]
 &=  \tilde{g}_{pq} - \frac{1}{2} \sum_r \left( \kappa_{pr} g_{rq} - \kappa_{qr} g_{rp} \right)  \label{eq:sig-orb-orb}
\end{align}
 of the regular ($\ham$) and the one-index transformed Hamiltonian
\begin{align}
 \htil &= \komu< \orb, \ham > = \sum_{pq} \kappa_{pq} \komu< \EOp_{pq}, \ham > \\
 &= \sum_{pq} \tilde{h}_{pq} \EOp_{pq} + \frac{1}{2} \sum_{pqrs} \widetilde{(pq|rs)} \E2Op_{pqrs}
\end{align}
Like $\ham$ the transformed Hamiltonian $\htil$ is also totally symmetric because only real totally symmetric orbital rotations are allowed
 for optimizing spin-restricted nonrelativistic wave functions.
As in our linear response implementation,\cite{Helmich-Paris2019} the integrals of $\htil$ are computed by a direct 
 transformation from the AO into the MO basis
\begin{align}
 \tilde{h}_{pq} &= \sum_{\mu\nu} \left( \Lambda_{\mu p} C_{\nu q} +  C_{\mu p} \Lambda_{\nu q} \right) \, h_{\mu\nu} \\
 \widetilde{(pq|rs)} &= (1+\hat{\mathcal{P}}_{pq,rs}) \sum_{\mu\nu\gk\gl} \left( \Lambda_{\mu p} C_{\nu q} +  C_{\mu p} \Lambda_{\nu q} \right) \notag \\
 & \phantom{(1+\hat{\mathcal{P}}_{pq,rs}) \sum_{\mu\nu\gk\gl}} C_{\gk r} C_{\gl s} \, (\mu\nu|\gk\gl)
\end{align}
 when using regular ($C_{\mu p}$) and trial vector-containing MO coefficients
\begin{align}
 \Lambda_{\mu p} =& - \sum_q C_{\mu q} \kappa_{qp}
\text{.}
\end{align}
For the latter, we exploit that there are only three nonredundant blocks in $\kappa_{pq}$ 
 (Eq.\ \eqref{eq:k-non-red}) with $p>q$ and that $\bkap$ is an antisymmetric matrix.
The same holds for the orbital gradient $g_{pq}$ contribution in Eq.\ \eqref{eq:sig-orb-orb}. 

The orbital gradient $\tilde{g}_{pq}$ of the one-index transformed Hamiltonian
 is computed by Eqs.\ \eqref{eq:g-ai} -- \eqref{eq:g-at} as $g_{pq}$,
 but instead, the integrals $\tilde{h}_{pq}$ and $\widetilde{(pq|rs)}$ are used.
In analogy to the orbital gradient,
 $\sigma_{pq}(\bkap)$ is computed by the corresponding tilded intermediates 
 $\widetilde{F}^I_{pq}$, $\widetilde{F}^A_{pq}$, and $\widetilde{Q}_{pt}$,
  which are compiled in Tab.\ \ref{tab:1}.

Similarly, the orbital-configuration sigma vector is formulated as 
an orbital gradient
\begin{align}
 & \sigma_{pq}( \bfS) = {\bfH}^{oc}\, {\bfS} \\[0.5em]
 &= - \sum_j w_j \left( \bra{ S_j} \left[ \EOp^-_{pq}, \ham \right] \ket{0_j} + \bra{ 0_j} \left[ \EOp^-_{pq}, \ham \right] \ket{S_j} \right) \\[0.5em]
 &=  \overline{g}_{pq}  \label{eq:sig-orb-cfg}
\end{align}
 for which the SA derivative density matrices $\overline{D}_{tu}$ and $\overline{d}_{tuvw}$
 and derivative overlap $\overline{\braket<0|S>} = 0$
 replace their analogues in Eqs.\ \eqref{eq:g-ai}--\eqref{eq:g-at}.
This additionally requires the computation of the $\overline{F}^A_{pq}$ and $\overline{Q}_{pt}$ intermediates
 which are given in Tab.\ \ref{tab:1}.

The configuration-orbital sigma vector is calculated in a similar fashion
 as the configuration gradient but employs the one-index transformed Hamiltonian
\begin{align}
  & \sigma_{Ii}( \bkap) = - 2 \, w_i \sum_{I'} \mathcal{U}_{II'} \, \bra{\Phi_{I'}} \htil \ket{ 0_i} \text{.} \label{eq:sig-cfg-orb}
\end{align}
Intermediates of the CAS-CI sigma vector computation in Eq.\ \eqref{eq:sig-cfg-orb}
  with $\htil$ are given in Tab.\ \ref{tab:1}.
Similarly, a configuration-configuration sigma vector is computed 
\begin{align}
   \sigma_{Ii}( \bfS ) = 2 \, w_i &\left( \sum_{I'} \mathcal{U}_{II'} \, \bra{\Phi_{I'}} \ham \ket{ \Phi_{J'} } \sum_J \mathcal{V}_{J'J} S_{Ji} \right. \notag \\[0.4em]
& \left. - \sum_j \bra{0_i} \ham \ket{0_j} S_{Ij} \right) \label{eq:sig-cfg-cfg}
\end{align}
that additionally demands a transformation of the configuration trial vectors
 from the nonredundant orthogonal-complement into the CSF basis by means of $\boldsymbol{\mathcal{V}}$,
before the CAS-CI sigma vector can be computed.
Note that, in general, the state-interaction Hamiltonian in Eq.\ \eqref{eq:sig-cfg-cfg}
 is not diagonal unless there are two or more states per spin
 or PG \textit{irrep} that have different weights $w_i$ as described in Sec.\ \ref{sec:egh}.

%--------------------------------------------------------------------%
\subsection{Basis transformation of configuration vectors} \label{sec:ci-bas}
%--------------------------------------------------------------------%

As worked out in the textbook Ref.\ \citenum{MEST-ci-bas},
 we make use of a consecutive chain of unitary operators to transform
 between the CSF and orthogonal complement basis back and forth.
For a single state, denoted $1$, such a unitary operator is given by
\begin{align}
 \hat{U}_1 &= 1 - \frac{ (\ket{0_1} - \ket{\Phi_{b(1)}})(\bra{0_1} - \bra{\Phi_{b(1)}})  }{1- \braket<0_1| \Phi_{b(1)}>}
\end{align}
 and rotates a selected, so-called \textit{basic}, CSF\cite{Olsen1985} $\ket{\Phi_{b(1)}}$
 onto the current solution $\ket{0_1}$  and all
 other CSFs into the orthogonal complement space.
For each state $i$, we choose the CSF with the largest-magnitude CI coefficient as \textit{basic CSF} $b(i)$.
For multiple states, consecutive application of the unitary operators
\begin{align}
 \hat{U}_i &= 1 - \frac{ \left( \widetilde{\ket{0_i}} - \ket{\Phi_{b(i)}} \right)
\left(\widetilde{\bra{0_i}} - \bra{\Phi_{b(i)}} \right)  }{1- \widetilde{\bra{0_i}} \Phi_{b(i)} \rangle } \\
 \widetilde{\ket{0_i}} &= \hat{U}_{i-1}\cdots\hat{U}_1\ket{0_i} \label{eq:proj-state}
\end{align}
rotates the CSFs either onto the set of $n$ current solutions $\ket{0_i}$
 or onto the orthogonal complement space of dimension $N-n$, that is,
\begin{align}
 \hat{U}_1 \cdots \hat{U}_{n} \ket{\Phi_K} &= 
\left\{
\begin{matrix}
\ket{0_k} \text{ if $K = b(k)$} \\[1.0em]
\ket{K} \text{ else }
\end{matrix}
\right.
 \text{.}
\end{align}
The equation above is abbreviated as
\begin{align}
 \hat{\mathcal{U}} \ket{\Phi_K} &= \ket{K} \text{.} \label{eq:csf2vec}
\end{align}
and occurs in its matrix representation (\textit{vide infra})
 in Eqs.\ \eqref{eq:g-cfg}, \eqref{eq:sig-cfg-orb}, and \eqref{eq:sig-cfg-cfg}.
Reversing the order of the state-specific unitary operators gives the transformation
 from the orthogonal complement to the CSF basis
\begin{align}
 \hat{U}_n \cdots \hat{U}_{1} \ket{K} &= \hat{\mathcal{V}} \ket{K} = \ket{\Phi_K} \text{,} \label{eq:vec2csf}
\end{align}
 which is required for $\sigma_{Ii}(\bfS)$ and for the derivative density matrices
  $\overline{D}_{tu}$ and $\overline{d}_{tuvw}$.

To find a more computationally amenable expression for the basis transformation, 
 the identity in CSF space $\sum_K \ket{ \Phi_K } \bra{ \Phi_K }$
 is inserted in Eqs.\ \eqref{eq:csf2vec} and \eqref{eq:vec2csf},
 which results in state-specific unitary matrices in CSF space
\begin{align}
  {\bfU}_i = \delta_{IJ} - \frac{ \left( \tilde{C}^i_I - \delta_{b(i),I} \right) \left( \tilde{C}^i_J - \delta_{b(i),J} \right) }
{1- \tilde{C}^i_{b(i)}} \label{eq:ci-trafo-mat}
\end{align}
and is the matrix representation of $\hat{U}_i$.
Concerning the computational effort of using Eq.\ \eqref{eq:ci-trafo-mat} for CI basis transformation,
  one should note that ${\bfU}_i$ are rank-one matrices and
 their matrix-vector products are computed only with $\mathcal{O}(n_{\text{SA}} \times N_{\text{CSF}})$ costs. 
Furthermore, the projected CI coefficients $\tilde{C}^i_I$ in Eq.\ \eqref{eq:ci-trafo-mat}
 are precomputed in every macro-iteration according to Eq.\ \eqref{eq:proj-state}
 and scale only with $\mathcal{O}(n^2_{\text{SA}} \times N_{\text{CSF}})$.
Thus, compared to other steps in the second-order CASSCF implementation, 
as the integral transformation or the CI sigma vector computation,
 the costs for the CI basis change are negligible.

%--------------------------------------------------------------------%
\subsection{Update of wave function parameters}
%--------------------------------------------------------------------%

Since an exponential parametrization is chosen for the variations in both
 the orbital and configuration space,
 the same approach based on eigenvalue decomposition of the anti-symmetric
 parameter matrix $\bfX$ is used.
First, the negated square of $\bfX$ is
 computed and then diagonalized\cite{MEST-orb-upd}
\begin{align}
 -\bfX \bfX = \bfW \btau^2 {\bfW}^T \label{eq:wf-up-eig}
\end{align}
The unitary wave function parameter update in its exponential form
 is then given by
 the eigenvalues and eigenvectors from Eq.\ \eqref{eq:wf-up-eig}
\begin{align}
 \exp(-\bfX) &= \bfW \cos(\btau) {\bfW}^T  \notag \\
 &- \bfW \btau^{-1}\sin(\btau) {\bfW}^T \, \bfX \label{eq:wf-up-gen}
\end{align}
To reach numerical stability for small values in $\btau$, 
 the truncated Taylor expansion is employed
\begin{align}
  \sin(\tau_i) / \tau_i \approx 1 - \frac{x^2}{3!} + \frac{x^4}{5!}
\end{align}
in those cases.

For updating the MO coefficients, Eq.\ \eqref{eq:wf-up-gen} can be readily used
 when assigning the orbital rotation matrices $\bkap$ with $\bfX^{o}$
\begin{align}
 \bfC &\leftarrow \bfC \exp( -\bfX^o ) \\[0.5em]
 \bfX^o &=  
\begin{pmatrix}
 \bNul      & - (\bkap_{ai})^T & - (\bkap_{vi})^T \\
 \bkap_{ai} &            \bNul & - (\bkap_{va})^T \\
 \bkap_{vi} &       \bkap_{va} & \bNul 
\end{pmatrix} \label{eq:orb-rot-block}
\end{align}
In Eq.\ \eqref{eq:orb-rot-block} $i$, $a$, and $v$ denote the inactive, active, and virtual MO subspace, respectively.

The update of the CI coefficients in exponential parametrization is less
 trivial because $\exp(\hat{S})$ rotates multiple states simultaneously
 in an SA calculation.
By inspecting the Taylor expansion of $\exp(\hat{S})$ term by term, it becomes obvious that
 the CI update can be expressed as
\begin{align}
\begin{pmatrix}
 \ket{ 0_1 } \\
 \ket{ 0_2 } \\
\vdots  \\
 \ket{ 0_n }
\end{pmatrix}
\leftarrow
 & \exp( -\hat{S})
\begin{pmatrix}
 \ket{ 0_1 } \\
 \ket{ 0_2 } \\
\vdots  \\
 \ket{ 0_n }
\end{pmatrix}
 =
 \cos( \bfX^{c} )
\begin{pmatrix}
 \ket{ 0_1 } \\
 \ket{ 0_2 } \\
\vdots  \\
 \ket{ 0_n }
\end{pmatrix}  \notag \\
 &- (\bfX^c)^{-1}
 \sin( \bfX^c )
\begin{pmatrix}
 \ket{ S_1 } \\
 \ket{ S_2 } \\
\vdots  \\
 \ket{ S_n }
\end{pmatrix}
\end{align}
with $\ket{ S^i }$ being the CI update vector of state $i$
 in the CSF basis and
\begin{align}
 X^c_{ij} = \left( \braket< S^i | S^j > \right)^{1/2}
\end{align}
 their norm matrix.
As for the orbital update, the cosine and sine terms are computed 
 from $\bfX^c$ according to Eqs.\ \eqref{eq:wf-up-eig} and \eqref{eq:wf-up-gen}.

%--------------------------------------------------------------------%
\section{Implementation Details} \label{sec:implement}
%--------------------------------------------------------------------%

\subsection{Choice of the orbital subspaces}

In previous CASSCF energy, second-order minimizer implementations\cite{Jensen1987}
 it was found that the convergence rate of the microiterations
 crucially depends on the particular form of
 the inactive, active, and virtual MOs.
In this work, we follow Jensen \textit{et al.}\cite{Jensen1987} and choose canonical inactive and virtual MOs.
This means that with those MOs the inactive-inactive and virtual-virtual blocks
 of the total MO Fock matrix become diagonal
\begin{align}
  F^I_{pq}+F^A_{pq} = \delta_{pq} \epsilon_p
\end{align}
 for $p,q = i,j$ or $p,q=a,b$, with the quasi-orbital energies $\epsilon_p$.

By default, in our implementation, the active MOs are rotated such
 that they keep the one-particle density matrix diagonal 
\begin{align}
 D_{tu} = \delta_{tu} n_t  \text{,}
\end{align}
which is referred to as natural active MOs\cite{Jensen1987} with $n_t$ being the natural occupation numbers.
Alternatively, localized active MOs can be used, which is customary
 when studying spin-coupled systems\cite{Fink2013,Chilkuri2019}
 and is further discussed in Sec.\ \ref{sec:spin_coupl}.

The current CI solutions are then counterrotated\cite{Malmqvist1986} if the active orbitals
 were rotated in the natural or localized active MO basis.

In case of SA calculations with unequal weights, a state-interaction
 basis for the CI coefficients must be chosen, which conflicts with the required 
 counterrotations due to an active MO basis change.
For such calculations, the active MOs remain unaltered.
After rotating MO coefficients of the three subspaces using in a particular basis,
 we need to recompute the MO Fock matrices and integrals before entering the micro-iterations.
The savings in the micro-iterations due to an MO basis change are usually much large than
 the extra costs of the recomputation.

%-----------------------------
\subsection{CAS-CI sigma vectors} \label{sec:impl-cas-ci}
%-----------------------------
The CAS-CI sigma vectors (Eq.\ \eqref{eq:ci-sigma}) and densities (Tab.\ \ref{tab:1}) are evaluated
with the bonded function method\cite{Manne1985} in combination with 
the second quantization based technique\cite{Golebiewski1985}
for the evaluation of matrix elements.\cite{Ganyushin2006}

%-----------------------------
\subsection{Integral transformation}
%-----------------------------

The computation of all gradient and sigma vector contributions
 in an integral-direct implementation
 requires two-electron integrals in the MO basis 
 with either three or four active indices.
To reduce the costs of the integral transformation from the AO
 to the MO basis, we exclusively employ the resolution-of-the-identity (RI)
 approximation with Coulomb metric for this intermediate.
The regular two-electron integrals with three active indices $(pu|vw)$
 are computed in every macro-iteration in the following way:
\begin{align}
 & B^Q_{pu} = \sum_{\mu} C_{\mu p} \sum_{\nu}  (Q|\mu\nu) C_{\nu u} \label{eq:ri-trafo} \\
 & V_{PQ} C^Q_{vw} = B^P_{vw} \label{eq:ri-solve} \\
 & (pu|vw) \approx \sum_{Q} B^Q_{pu} C^Q_{vw} \label{eq:ri-contr} 
\end{align}
In the equations above, $P$ and $Q$ denote auxiliary AO basis functions.
The all-active integrals needed for the CAS-CI vectors are included in $(pu|vw)$
To tackle linear dependencies in the auxiliary basis set and to ensure numerical stability when solving
 Eq.\ \eqref{eq:ri-solve}, the pivoted Cholesky factors of $V_{PQ}$ are employed
 that were obtained before the CASSCF calculation.

In every micro-iteration, the two-electron integrals of the one-index
 transformed Hamiltonian must be computed additionally for which we also use RI as was shown in Ref.\ \citenum{Helmich-Paris2019}
\begin{align}
 & \widetilde{B}^Q_{pu} = \sum_{\mu\nu } (Q|\mu\nu) \left( \Lambda_{\mu p} C_{\nu u} + C_{\mu p} \Lambda_{\nu u} \right) \label{eq:ri-rsp-trafo} \\
 & \sum_{Q} V_{PQ} \widetilde{C}^Q_{vw} = \widetilde{B}^P_{vw} \label{eq:ri-rsp-solve} \\
 & \widetilde{(pu|vw)} \approx \sum_{Q} \left( \tilde{B}^Q_{pu} C^Q_{vw} + B^Q_{pu} \tilde{C}^Q_{vw} \right) \label{eq:ri-rsp-contr} 
\text{.}
\end{align}

Note that for very large molecules and medium sized active space sizes,
 Eq.\ \eqref{eq:ri-trafo} is the most time-consuming step in 
 our Super-CI (SX) PT-\cite{Kollmar2019} and TRAH-CASSCF implementation
 that scales with $\mathcal{O}( N_{\text{aux}}\, N^2 \, N_{\text{act}})$.
Since the number of active orbitals $N_{\text{act}}$ is always limited by the steep scaling of the 
 CAS-CI method to a small number and since the number of auxiliary basis functions is usually
 three to five times larger than the number of orbital basis functions, the overall scaling
 of the integral transformation
 is $\mathcal{O}(\mathcal{N}^3)$ with the system size $\mathcal{N}$.

%-----------------------------
\subsection{Fock matrices} \label{sec:impl_fock}
%-----------------------------

The other bottleneck of our CASSCF implementations, when choosing moderately large active spaces,
 is the computation of the AO Fock matrices.
In every macro-iteration, $F^I_{\mu\nu}$ and $F^A_{\mu\nu}$ must be computed,
 in every micro-iteration $2+n_{\text{SA}}$ AO Fock matrices $G_{\mu\nu}[\bfD]$ with $\bfD$ being
 either the core $\widetilde{\bfD}^I$ or active $\widetilde{\bfD}^A$
 orbital derivative
 or the configuration derivative $\overline{\bfD}^A$ AO density matrices.
Besides computing Fock matrices from the usual four-center fully analytic two-electron integrals,
 we have also implemented two approximate, though much more efficient algorithms.
For small- and medium-sized molecules, we employ the RI approximation for
 both the Coulomb (J)
 and exchange (K) term
 known as RIJK.\cite{Kendall1997,Kossmann2009}
The RI-J matrices are still computed entirely in the AO basis\cite{Whitten1973,*Baerends1973,*Dunlap1977,*Vahtras1993,*Eichkorn1995} using the split algorithm.\cite{Neese2003}
Thus, a single RI-J algorithm can be used for all the five types of AO density matrices (\textit{vide supra})
\begin{align}
 & \rho_P = \sum_{\gk\gl} (P|\gk\gl) D_{\gk\gl} \\
 & \sum_Q V_{PQ}\, j_Q = \rho_P \\
 & J_{\mu\nu}[\bfD] = \sum_{Q} (\mu\nu|Q) j_Q \text{.}
\end{align}
Conversely, the RI-K matrices
 require a dedicated algorithm for each type
\begin{align}
  K_{\mu\nu}[\bfD^I] &=  2\, \sum_{Qi} B^Q_{\mu i} C^Q_{i \nu } \label{eq:ri-k-i} \\
  K_{\mu\nu}[\bfD^A] &=      \sum_{Qtu} B^Q_{\mu u} D_{tu} C^Q_{t \nu} \label{eq:ri-k-a} \\
  K_{\mu\nu}[\widetilde{\bfD}^I] &=  2\, \sum_{Qi} \left( \tilde{B}^Q_{\mu i} C^Q_{i \nu} + B^Q_{\mu i} \tilde{C}^Q_{i \nu} \right) \label{eq:ri-k-orb-i} \\
  K_{\mu\nu}[\widetilde{\bfD}^A] &=  \sum_{Qtu} \left( \tilde{B}^Q_{\mu u} D_{tu} C^Q_{t \nu} + B^Q_{\mu u} D_{tu} \tilde{C}^Q_{t \nu} \right) \label{eq:ri-k-orb-a} \\
  K_{\mu\nu}[\overline{\bfD}^A]  &=  \sum_{Qtu} B^Q_{\mu u} \overline{D}_{tu} C^Q_{t \nu} \label{eq:ri-k-cfg-a}
\end{align}
 because they are computed from partially transformed inactive or active MO intermediates
\begin{align}
B^Q_{\mu p} &= \sum_{\nu} (Q|\mu\nu) C_{\nu p} \\
\tilde{B}^Q_{\mu p} &= \sum_{\nu} (Q|\mu\nu) \Lambda_{\nu p} \text{,}
\end{align}
 and $C^Q_{\mu p}$ and $\tilde{C}^Q_{\mu p}$ with $p = i$ (inactive) or $p = t,u$ (active).

While the RI-J matrix computation scales for large systems as $\mathcal{O}(\mathcal{N}^2)$ if Schwartz screening is employed,.\cite{Almloef1982,Haeser1989}
 the two major RI-K steps scale both with $\mathcal{O}(\mathcal{N}^4)$ and limit the approach to medium sized systems.
Due to its small prefactor, the RIJK method is very efficient for smaller molecules\cite{Kossmann2009}
 as will be demonstrated in Sec.\ \ref{sec:results}.
We note that it would be even more efficient to compute the exchange matrix 
 contribution to the four relevant orbital gradient blocks directly in the MO basis.
However, this has not been attempted in this work because such an MO-based implementation
 would differ more significantly from our current AO-based implemented.

% COSX
For large systems, the $\mathcal{O}(\mathcal{N}^4)$ scaling and I/O operations of the RI-K method
 becomes a bottleneck.
Therefore, the latest version of the $\mathcal{O}(\mathcal{N})$ scaling 
 chain-of-spheres for exchange algorithm\cite{Helmich-Paris2021,Neese2009,Izsak2011} (COSX)
 is employed for such calculations.
This type of semi-numerical exchange algorithms
\begin{align}
K_{\mu\gl}  &=  \frac{1}{2} \left( \mathcal{K}_{\mu\gl} + \mathcal{K}_{\gl\mu} \right)  \\
 \mathcal{K}_{\mu\gl} &=  \sum_{\gk\nu} \sum_g \left( X_{\mu g} X_{\nu g} A^g_{\gk \gl} \right) D_{\gk\nu} \\
 A^g_{\gk\gl} &= \int d\bfr \, \chi_{\gk}(\bfr) \, \chi_{\gl}(\bfr) \, \frac{1}{| \bfr - \bfr_g |} \\
 X_{\mu g} &= |w_g|^{1/2} \, \chi_{\mu}(\bfr_g)
\end{align}
 performs one integration
  numerically by a weighted ($w_g$) sum over grid points $\bfr_g$ and the other analytically 
  by means of electrostatic potential integrals  $\bfA^g$ located at $\bfr_g$.
In principle, no further adjustments 
 or additions of the COSX algorithm are necessary because it works 
 entirely in the AO basis and multiple AO density matrices
 can be processed simultaneously,
 that is, $\bfD^I$ and $\bfD^A$ for the macro-iterations and 
 $\widetilde{\bfD}^I$, $\widetilde{\bfD}^A$, and $\widetilde{\bfD}^A$
 for the micro-iterations.
However, we realized that the loss of the eight-fold permutational symmetry 
 of the two-electron integrals introduced by overlap (S)- fitting,\cite{Izsak2011} 
\begin{align}
 \mathcal{K}_{\mu\gl} &= \sum_{\gk\nu} \sum_g \left( Q_{\mu g} X_{\nu g} A^g_{\gk \gl} \right) D_{\gk\nu} \\
 \bfQ &= ( \bfS (\bfX \, {\bfX}^T )^{-1} ) \bfX \text{,}
\end{align}
leads to noticeable numerical errors and convergence issues,
 in particular, for calculations with transition metal-containing molecules. 
In those calculations, we used an S-fitted COSX variant that preserves the full
 eight-fold permutational symmetry
\begin{align}
 \mathcal{K}_{\mu\gl} = \frac{1}{2} \sum_{\gk\nu} \sum_g  
  &\left( X_{\mu g} Q_{\nu g} A^g_{\gk \gl} \right. \notag \\
+ &\left. Q_{\mu g} X_{\nu g} A^g_{\gk \gl} \right) D_{\gk\nu}
\end{align}
 and drastically reduces the numerical errors.
Thus, the symmetrized S-fitted COSX variant eases convergence
 at the expense of slightly larger costs.\cite{BHPFantasy}

In our TRAH implementation,
 the many costly exchange matrices in the micro-iterations
 are evaluated with a smaller grid than in the macro-iterations.
This automatic grid reduction has already turned out to be a reasonable approximations
 when solving coupled-perturbed SCF equations,\cite{Petrenko2011}
 which significantly reduces the costs of AO Fock matrices without introducing
 significant errors.
By default, for numerical integration in COSX, 
 we employ a 50-point Lebedev angular grid
 and a radial grid with an integration accuracy parameter\cite{Krack1998} $\varepsilon = 3.067$
 during the micro-iterations.

%-----------------------------
\subsection{Preconditioning}
%-----------------------------

To obtain fast convergence of the micro-iterations,
 it is important to choose a preconditioner  $\bfM$
 that is a good approximation to
\begin{align}
 & \left( \bfH - \mu \bfI \right)^{-1}
\end{align}
 but remains computationally feasible.
For $\bfH$ in $\bfM$, we only include the
  $\bfH^{oo}$ and  $\bfH^{cc}$ blocks.
For the majority of $\bfH^{oo}$ matrix elements in $\bfM$,
 we only take the approximate diagonal Hessian\cite{Chaban1997,Helmich-Paris2019}
 $\bfD^{oo}$
 which can be quickly computed from the $\bfg^{o}$ intermediates.
But a small subblock of $\bfH^{oo}$ is still computed explicitly.
A subset of nonredundant orbital pairs, which we now refer to as \textit{reduced}, is selected in the following way:
 (i) All active MOs are considered.
 (ii) We select a few virtual-inactive orbital pairs
 with the smallest pseudo-orbital energy difference $\epsilon_a - \epsilon_i$ ( 250 by default).
 The smallest inactive and the largest virtual MO index in that list
  defines the range of the subset MOs
 together with all active MOs.
All nonredundant MO pairs from that MO subset define the \textit{reduced} orbital pair list.
Then, the $\bfH^{oo}$ subblock\cite{Siegbahn1980} of all orbital pairs from that list
 is computed by using the RI approximation for all types of two-integral integrals that occur in $\bfH^{oo}$.
Thereafter,
 the $\bfH^{oo}$ subblock is diagonalized
\begin{align}
& \bfH^{oo} \, \bfU^o  = \boldsymbol{\eta}^o \, \bfU^o  \text{.}
\end{align}
In the micro-iterations, orbital pairs of the residual vectors for $\bkap$ (see Ref.\ \citenum{Helmich-Paris2021})
 are then preconditioned either with $\bfH^{oo}$ eigenvalues and eigenvectors
\begin{align}
& \bfM^{oo} = \bfU^o \left( \boldsymbol{\eta}^o - \mu \bfI \right)^{-1} ( \bfU^o)^T
\end{align}
 if they are in the \textit{reduced} list
 or with $\bfD^{oo}$ 
\begin{align}
& \bfM^{oo} = \left( \bfD^{oo} - \mu \bfI \right)^{-1}
\end{align}
 if they are not.

For the $\bfH^{cc}$ block in $\bfM$, we proceed in a similar way.
The diagonal elements of the full CAS-CI matrix 
\begin{align}
\mathcal{H}_{IJ} &= \bra{\Phi_I} \ham \ket{\Phi_J} \label{eq:ci-mat}
\end{align}
are computed in the CSF basis and used
 for the diagonal approximation to $\bfH^{cc}$
\begin{align}
 \bfD^{cc} &= 2\, w_i \left( \bra{\Phi_I} \ham \ket{\Phi_I} - E_i \right)
\end{align}
To simplify the inversion, we have neglected all rank-one terms of the CSF based
 $\bfH^{cc}$ in $\bfM$.
A few of the smallest diagonal elements $\mathcal{H}_{II}$ (250 by default)
 define the CSFs $\{ \Phi_I \}$ in the \textit{reduced} list for the configuration
 parameter space.
For \textit{reduced}-list CSFs,
 the full CAS-CI matrix (Eq.\ \eqref{eq:ci-mat}) is computed 
 as briefly explained in Sec.\ \ref{sec:impl-cas-ci} and then diagonalized
\begin{align}
 \boldsymbol{\mathcal{H}}  \, \bfU^c  = \boldsymbol{\eta}^c \, \bfU^c  \text{.}
\end{align}
Before the configuration part of the residual vectors for $\bfS$
 can be preconditioned, they must be transformed from nonredundant, orthogonal-complement
 basis into the redundant CSF basis.
Then, in the micro-iterations the preconditioner with the $\boldsymbol{\mathcal{H}}$ eigenvalues and eigenvectors
\begin{align}
& \bfM^{cc} = \bfU^c \left( 2\, \delta_{ij}\, w_i \left( \boldsymbol{\eta}^c - E_i\bfI \right) - \mu \bfI \right)^{-1} ( \bfU^c)^T
\end{align}
 is used if the CSFs are in the \textit{reduced} list,
the diagonal preconditioner
 $\bfD^{cc}$  is used if they are not.

A similar preconditioning algorithm is used by Kreplin \text{et al.}\ for 
 their second-order (SO) SX CASSCF method.

%--------------------------------------------------------------------%
\section{Computational Details} \label{sec:compdet}
%--------------------------------------------------------------------%

The TRAH-CASSCF for SS and SA calculations was implemented in a development version of ORCA\cite{Neese2012}.
All SXPT-CASSCF calculations\cite{Kollmar2019} were performed with the same development version.
TRAH-CASSCF convergence is exclusive checked using the gradient norm $||\bfg||$.
If $||\bfg|| < 10^{-6}$, convergence is reached for TRAH.
By default, SXPT convergence is more loosely because orbital gradient norm $||{\bfg}^o||$ and 
 the magnitude of the energy difference between two consecutive iterations $|\delta E|$ is checked.
If only one criterion is fulfilled, SXPT signals convergence.
For SXPT, we chose the thresholds $10^{-5}$ and $10^{-10}$ for $||{\bfg}^o||$ and $|\delta E|$, respectively.

% always use RI for 2e integrals
We employ exclusively the RI approximation for the transformation
 of two-electron integrals with three and four active orbitals.
A global integral-neglect threshold for Schwartz screening\cite{Haeser1989} of $10^{-14}$ is used.
The latter is employed for Fock matrix construction and the RI integral transformation.
For the latter, a linear dependency threshold of $10^{-13}$ for the pivoted Cholesky decomposition was used.
%BFCut -12

% symmetrized COSX
All CASSCF calculations with COSX used the new tight default grid\cite{Helmich-Paris2021b} (\texttt{DefGrid3})
 for numerical integration.
Also, the fully symmetrized version of the S-fitted COSX was employed (see Sec.\ \ref{sec:impl_fock}).

Point-group symmetry was never exploited for any calculation.
For localizing the active orbitals the Foster-Boys procedure was used.\cite{Foster1960}

% GUESS: 
%  - AVAS procedure
For TMC calculations,
 we used exclusively the atomic-valence active-space (AVAS) procedure of Sayfutyarova et al.\cite{Sayfutyarova2017},
 which simultaneously provides both an MO guess and an active-space definition
 that can be represented by a user-given set of target atomic orbitals.
In our case, these are five 3d AOs for each TM center that we take from
 the minimal basis set MINAO\cite{Sayfutyarova2017}.
Guesses for double d-shell calculations were made available, too,
 by following the discussion in Ref.\ \onlinecite{Sayfutyarova2017}.
For the special case of minimal active-space or double d-shell calculations
 of TMCs, the only strictly necessary input quantity is the list of TM center
 indices in the molecules.

%  - UNO (RIJCOSX / UHF calculation + stability analysis + BS UHF
%  - PiAS (simplified version of PiOS)
For calculations on aromatic systems, we employed two different guesses for the initial MOs.
The first guess type are unrestricted natural orbitals\cite{Pulay1988} (UNO) that are
 obtained from a preceding broken-symmetry (BS) unrestricted Hartree--Fock (UHF) calculation.\cite{Nottoli2021}
BS-UHF solutions become available after an SCF stability analysis\cite{Seeger1976,*Seeger1977}
 and subsequent UHF calculation. 
The second guess is what we refer to as \texttt{PiAS} in the following and
 is a simplified version of the $\pi$ MO guess of Sayfutyarova and Hammes-Schiffer.\cite{Sayfutyarova2019}
In accordance with their work,\cite{Sayfutyarova2019} we define a tensor of inertia with unit masses
 for each center that is part of the desired $\pi$ system.
After choosing the principle axis with the largest moment, a linear combination of the three $2p$ target AO
 components multiplied by these particular axis coordinates 
 is processed when computing the AVAS MO projection matrix.\cite{Sayfutyarova2017}
The remaining steps coincide with AVAS.\cite{Sayfutyarova2017}
Hence, the \texttt{PiAS} procedure is capable of providing a reasonable MO guess for arbitrarily spatially oriented
 molecules and their target $\pi$ chromophores.

% structures
Most structures were taken from the respective articles and their Supplementary Material section. 
The structure of the dicobalt(II) oxo complex was kindly provided by the authors of Ref.\ \onlinecite{Roy2019}.
 
% basis sets
For the benchmark study on small aromatic molecules, we employed the cc-pVTZ basis sets\cite{Dunning1989,*Woon1993}
 and the corresponding auxiliary /JK basis sets\cite{Weigend2002b} for the RIJK approximation of Fock matrix and
 auxiliary /C basis sets\cite{Weigend2002} for the integral transformation.
The same basis sets were used for the chlorophyll calculations.
We have used the def2-TZVPP basis sets\cite{Weigend2005} and the corresponding def2/JK\cite{Weigend2008}
 and /C auxiliary basis sets\cite{Weigend1998,Hellweg2007}
 for the RI approximation in all other calculations.

%--------------------------------------------------------------------%
\section{Results and Discussion} \label{sec:results}
%--------------------------------------------------------------------%

%--------------------------------------------------------------------%
\subsection{Small aromatic systems}\label{sec:res_arom}
%--------------------------------------------------------------------%

\begin{figure}
 \centering
\begin{subfigure}{\linewidth}
 \centering
 \captionsetup{justification=centering}
 \begin{subfigure}{.48\textwidth}
  \centering
  \includegraphics[width=0.98\textwidth]{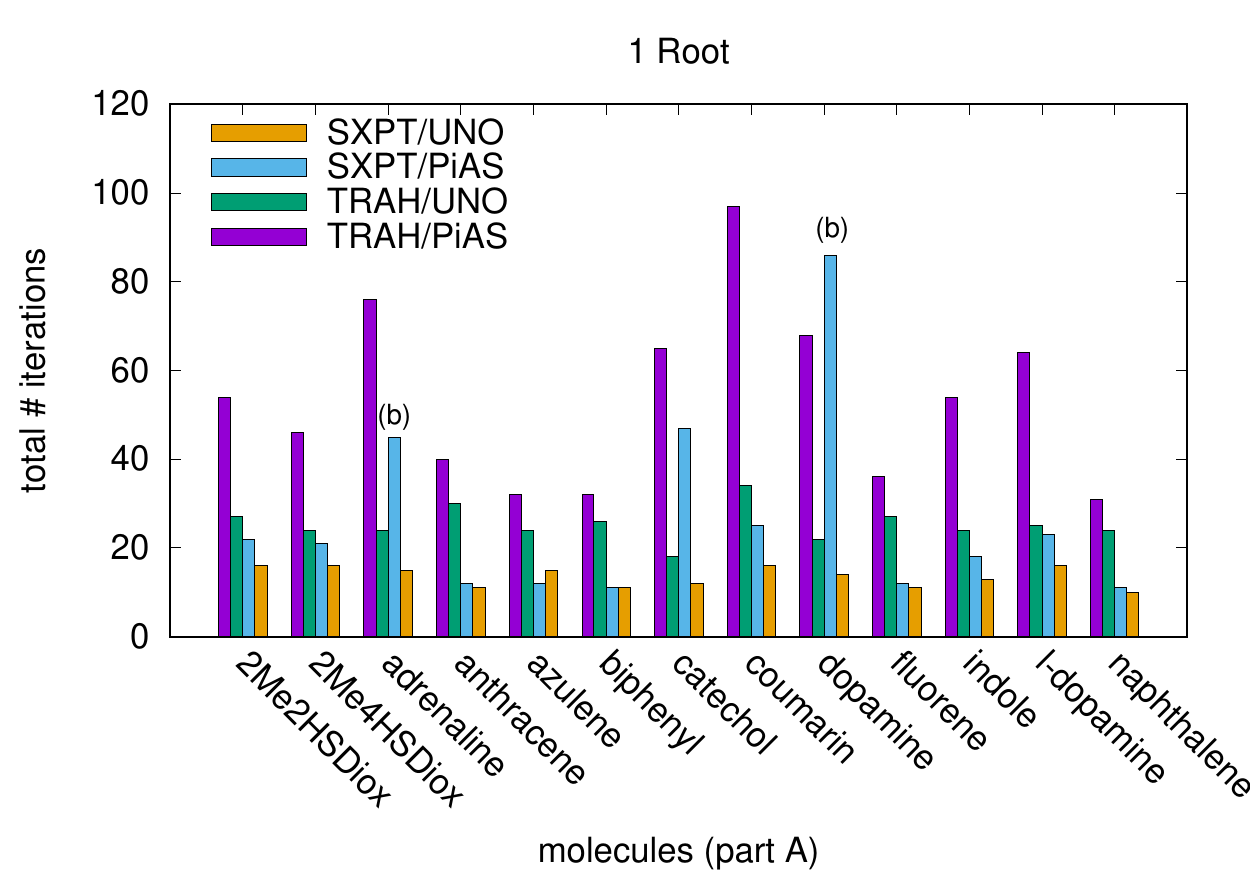}
  \caption{}
  \label{fig:3a}
 \end{subfigure}
 \begin{subfigure}{.48\textwidth}
  \centering
  \includegraphics[width=0.98\textwidth]{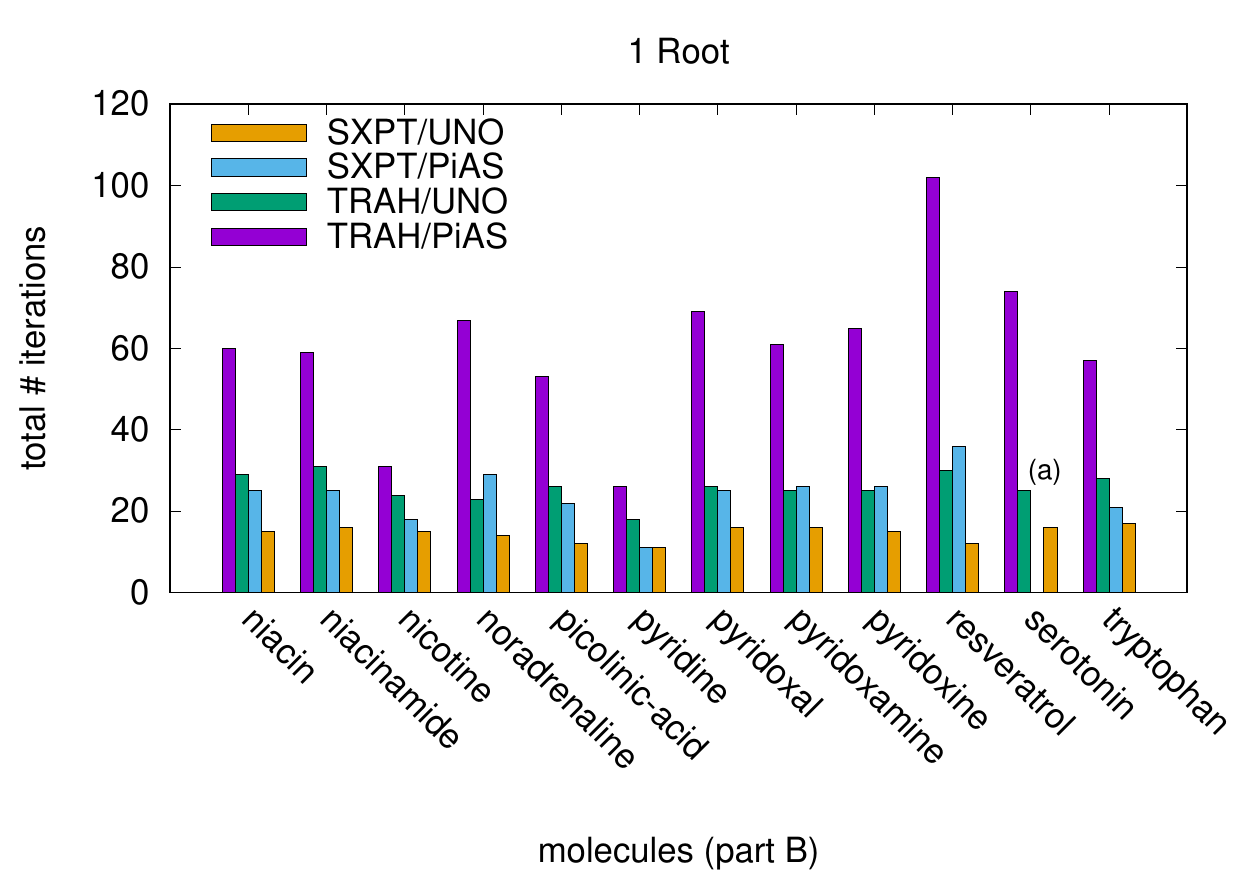}
  \caption{}
  \label{fig:3b}
 \end{subfigure}
 \begin{subfigure}{.48\textwidth}
  \centering
  \includegraphics[width=0.98\textwidth]{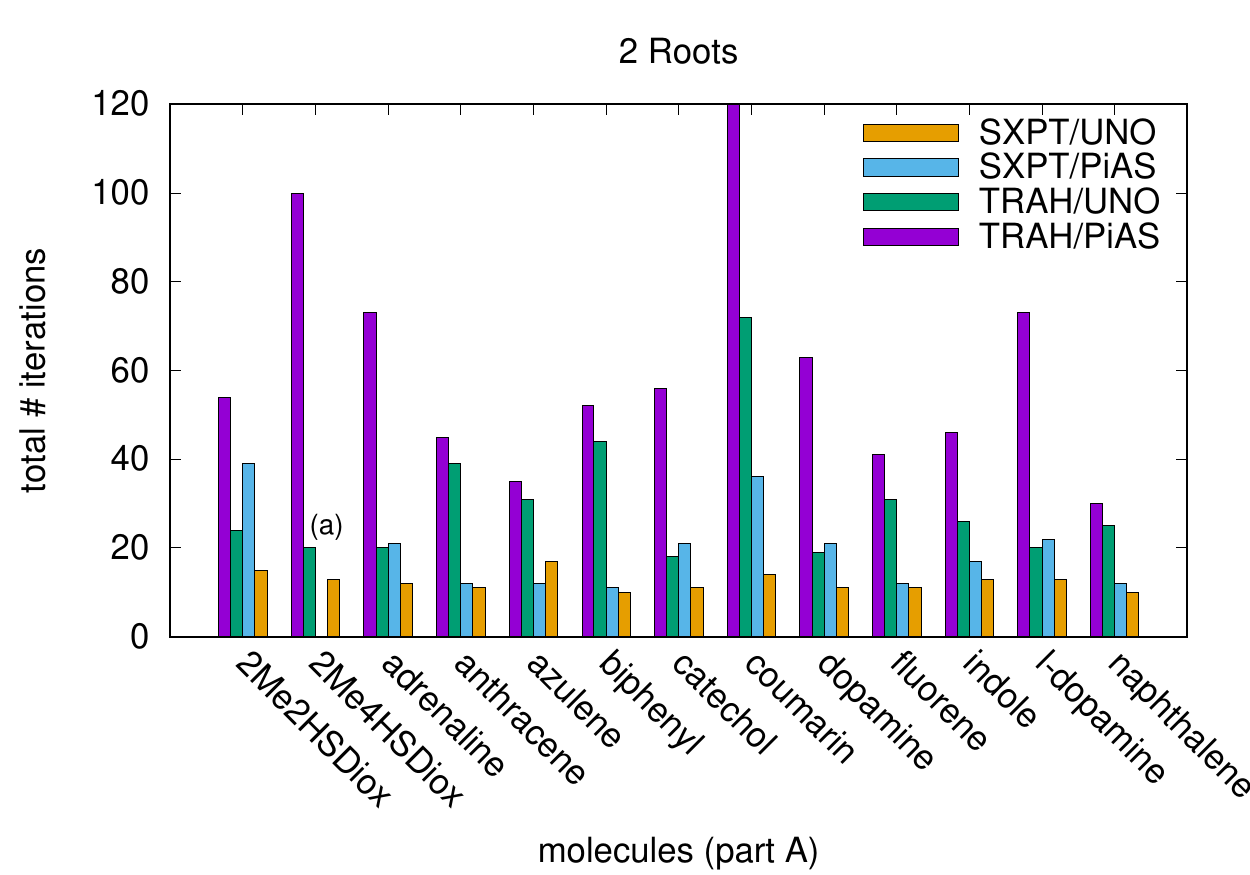}
  \caption{}
  \label{fig:3c}
 \end{subfigure}
 \begin{subfigure}{.48\textwidth}
  \centering
  \includegraphics[width=0.98\textwidth]{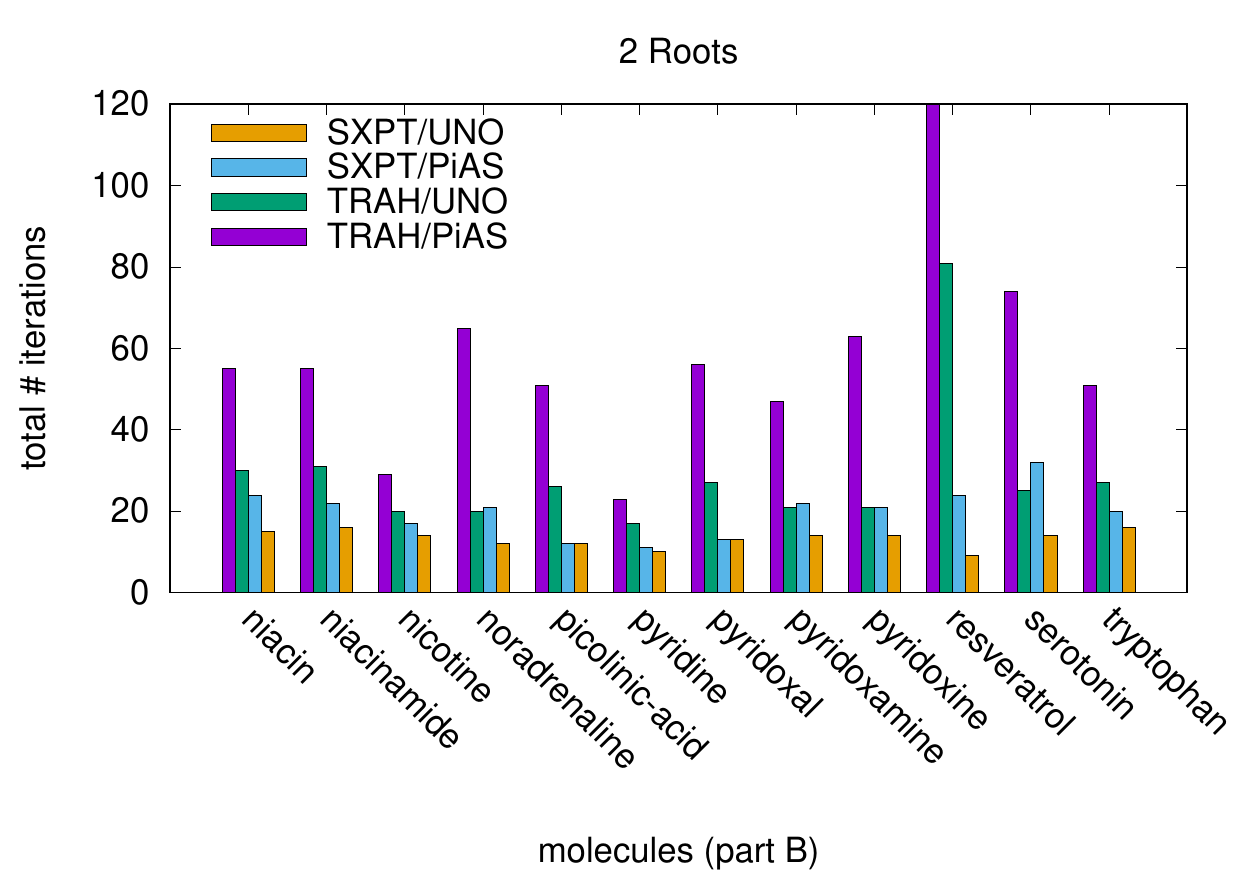}
  \caption{}
  \label{fig:3d}
 \end{subfigure}
	\end{subfigure}
 \caption{
Total number of iterations to converge the CASSCF/cc-pVTZ energy
 for either one root ((\subref{fig:3a}) and (\subref{fig:3b}) )
 or two roots ((\subref{fig:3c}) and (\subref{fig:3d})).
 Two different algorithms (SXPT or TRAH) and
  two different types of initial MOs (PiAS and UNO from BS-UHF solution)
  are used.
 The (a) and (b) labels inside the figures denote no convergence and convergence to a higher-energy
 solution, respectively.
}
 \label{fig:3}
\end{figure}

The first class of systems for which we compare the number of iterations and
 timings of our new TRAH-CASSCF implementation with other algorithms and implementations
 are small aromatic systems.
A benchmark set for MR methods was recently proposed by
 Menezes et al.\cite{Menezes2016} and then extended by Nottoli et al.\cite{Nottoli2021},
 which eventually comprises 25 small to medium-sized molecules.
To compare our results with those of the CD-NEO-CASSCF implementation of Nottoli et al.,
 we chose the same basis set (cc-pVTZ)
 and the same active spaces (See Tab.\ \ref{tab:2}) as they did.
We employed the RIJK approximation for the Fock matrices in those calculations
 as it features similar intermediates as the CD-NEO-CASSCF implementation and has the
 same scaling with the system size.
The total number of iterations for SXPT and TRAH are given in Fig.\ \ref{fig:3}
 for both SS and two-root SA calculations.
Erratic convergence behavior is also labeled in  Fig.\ \ref{fig:3}.
There, we see that
 SS- and SA-TRAH calculations always converge irrespective of the 
 choice of initial MOs.
With our TRAH implementation, the same minimum energy and solution is obtained
 when starting from the two different types of initial MOs and the final active MOs are the
 desired $\pi$ orbitals of the aromatic system.
However, this does not hold for the first-order SXPT.
Two SXPT calculations converged to a solution with a higher energy than
 TRAH, i.e.\ the single-root adrenaline and dopamine calculations both 
  starting with the PiAS guess. 
Two other SXPT calculations even diverged, i.e.\ single-root 
 serotonin calculation and the two-root 2Me4HSDiox calculation,
 which also started from the PiAS guess.
Furthermore, it can be seen from Fig.\ \ref{fig:3} that, for a particular MO guess,
 the first-order SXPT
 method always takes less iterations than TRAH if the number of macro and
 micro-iterations are counted together as it is done here and in the following.
Only in case of erratic convergence behavior, SXPT took more iterations than TRAH.
However, fewer iterations of SXPT does not necessarily correspond to a faster
 runtime of SXPT in comparison with TRAH, as can be seen in Tab.\ \ref{tab:2}.
For the largest active spaces in the benchmark set, CAS(14,14),
 TRAH is noticeably faster though it takes more iterations than SXPT.
This is caused by the many, costly CAS-CI vector computations in SXPT that are a consequence of
 the two-step algorithm which solves CAS-CI eigenvalues equations in every (macro) iteration.
A genuine one-step second-order implementation like TRAH improves the 
 CI solution basically in every iteration, in particular, when close to the minimum solution.

In Tab. \ref{tab:2}, we have also compiled the total runtime of the CD-NEO-CASSCF
 implementation, which was for this benchmark set always faster then TRAH,
 roughly by a factor 4 --5 for the smallest systems and by 1 -- 2 for the larger systems.
For a few cases, the first-order SXPT calculations were slightly faster than CD-NEO-CASSCF
 (2Me4HSDiox, azulene, and coumarin).

It should be noted here that a truly reliable comparison of timings cannot be
 made here because
 (i) we ran our SXPT and TRAH calculations with fewer number of processes (20)
  than the number of threads (28) used for CD-NEO-CASSCF in Ref.\ \onlinecite{Nottoli2021}.
 (ii) a different computing architecture was used, and
 (iii) subtle difference of the RIJK and CD approximation.

Concerning the efficiency, our new TRAH-CASSCF implementation is competitive
 with the first-order SXPT and another second-order implementation when using comparable
 algorithms for the integrals.
An efficiency comparison will also be made for larger molecules (\textit{vide infra}),
 for which efficiency has a higher significance. 

% state averaging for organic chromophores (TODO???)
%  - one-step vs two-step

%--------------------------------------------------------------------%
\subsection{Spin-coupled systems}\label{sec:spin_coupl}
\begin{figure}
 \centering
 \begin{subfigure}{\linewidth}
  \centering
  \captionsetup{justification=centering}
  \begin{subfigure}{.48\textwidth}
   \centering
   \includegraphics[width=0.98\textwidth]{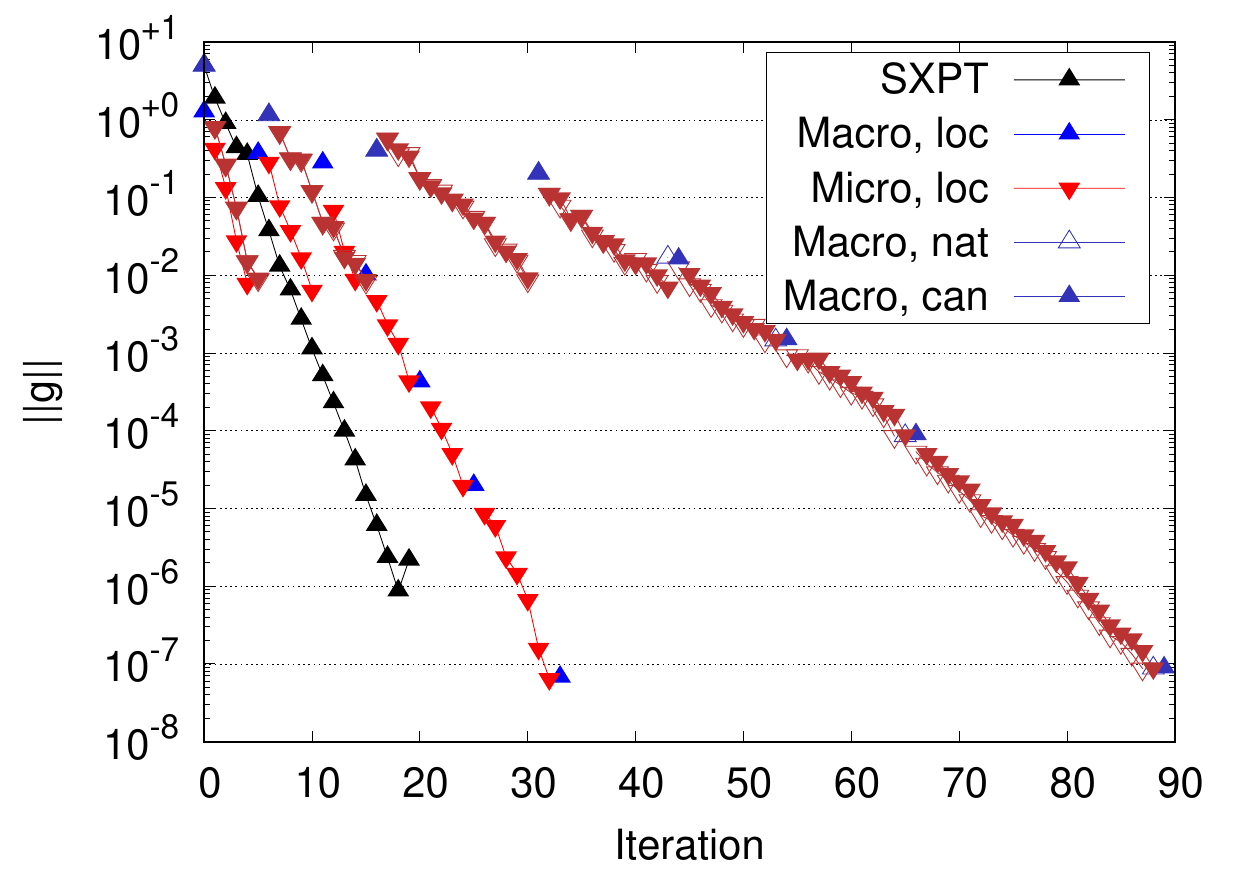}
   \caption{}
   \label{fig:4a}
  \end{subfigure}
  \begin{subfigure}{.48\textwidth}
   \centering
   \includegraphics[width=0.98\textwidth]{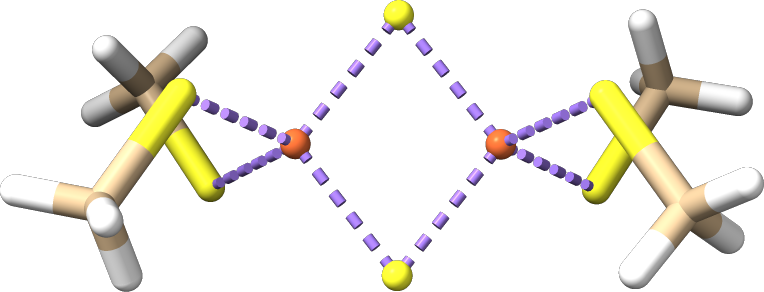}
   \caption{}
   \label{fig:4b}
	\end{subfigure}
 \end{subfigure}
 %\caption{.}
 \caption{
SXPT and TRAH convergence of CASSCF/def2-TZVPP for the ${^1}A_g$ state (\subref{fig:4a})
 of an iron-sulfur dimer molecule ($C_{2h}$ point group) (\subref{fig:4b}).
For TRAH calculations different active orbital choices were made:
 natural, canonical, and localized.
}
 \label{fig:4}
\end{figure}

%--------------------------------------------------------------------%
Another typical field of application for MR methods are spin-coupled systems.
Here we study the SS-CASSCF energy convergence of an anti-ferromagnetically
 coupled iron- (III) dimer complex\cite{Chilkuri2019} ( [ Fe\tief{2}S\tief{2}(SCH\tief{3})\tief{4}]\hoch{2-} )
 in the ${^1}A_g$ state.
The convergence of the gradient norm for the SXPT and TRAH
 CASSCF/def2-TZVPP calculations with natural, canonical, or localized active orbitals
 is shown in Fig.\ \ref{fig:4}.
A minimal active space CAS(10,10) that includes all Fe(III) 3d electrons and
 orbitals in the active space was chosen.
Initial orbitals were obtained from the model one-electron Hamiltonian
 and the AVAS procedure.
The SXPT algorithm converges steadily within 19 iterations,
 which shows that the initial MOs are well suited for these type of calculations
 and are coherent with the chosen AS.
The convergence rate of our new TRAH implementation depends crucially on the
 active orbital choice.
Canonical and natural active orbitals converge slowly and almost need 90 
 iterations.
However, localized active orbitals dramatically improve the convergence rate
 of the TRAH calculation, which can be attributed 
 to the disentanglement of the CAS-CI Hamiltonian into weakly coupled Fe(III) monomer blocks.
Each of these monomers is qualitatively described by a single high-spin
 CSF that are then antiferromagnetically coupled for the dimer.
This is confirmed by the fact that the antiferromagnetically monomer CSF leads the final CI wave function 
 expansion with a weight of 98~\%.
When employing canonical or natural active orbitals, we still converge to the same energy
 as with SXPT and with the localized active orbital TRAH calculation.
However, with those orbital choices, there are plenty of CSFs
 leading the final CI wave function expansion with weights less than 0.02~\%.

%--------------------------------------------------------------------%
\subsection{ Degenerate and nearly degenerate ground states}
\begin{figure}
 \centering
 \begin{subfigure}{\linewidth}
  \centering
  \captionsetup{justification=centering}
  \begin{subfigure}{.48\textwidth}
   \centering
   \includegraphics[width=0.98\textwidth]{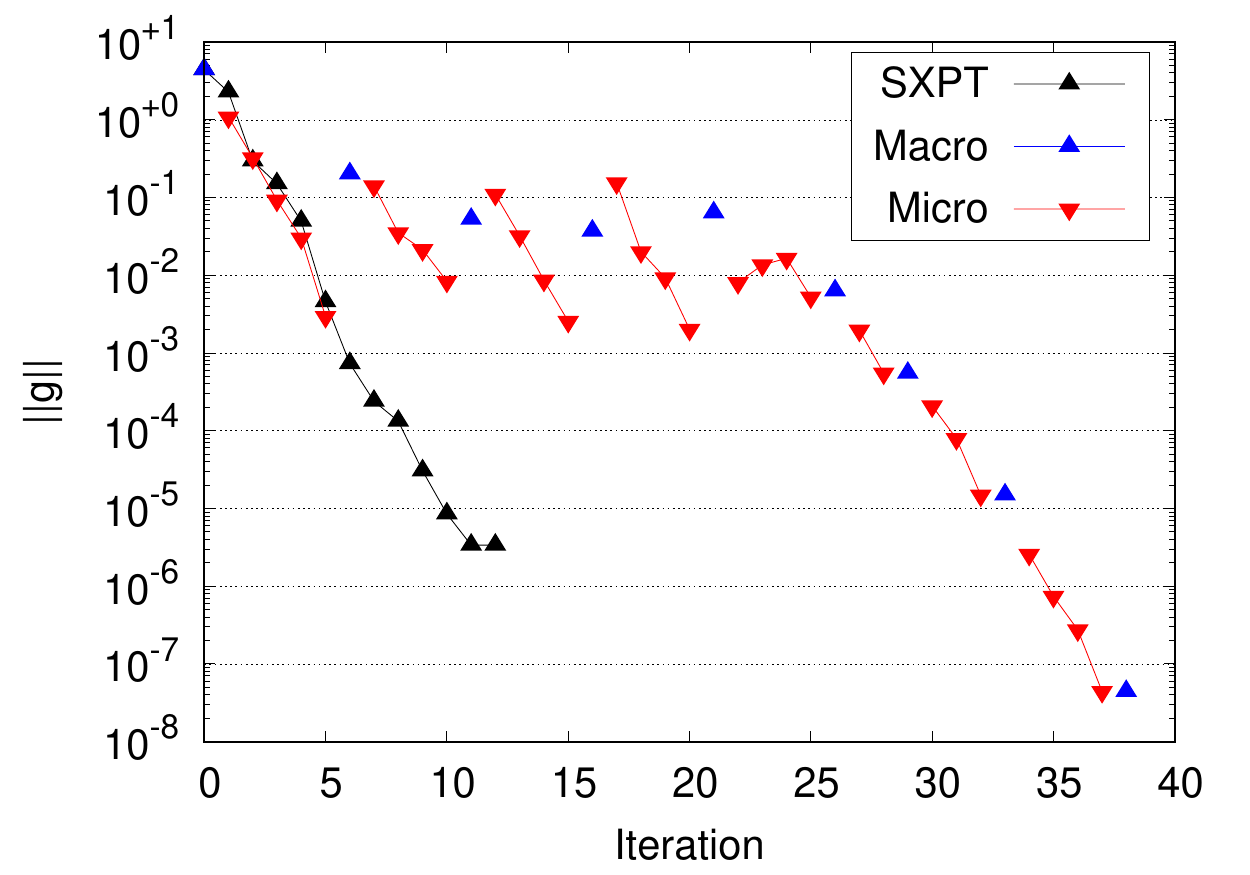}
   \caption{}
   \label{fig:5a}
  \end{subfigure}
  \begin{subfigure}{.48\textwidth}
   \centering
   \includegraphics[width=0.98\textwidth]{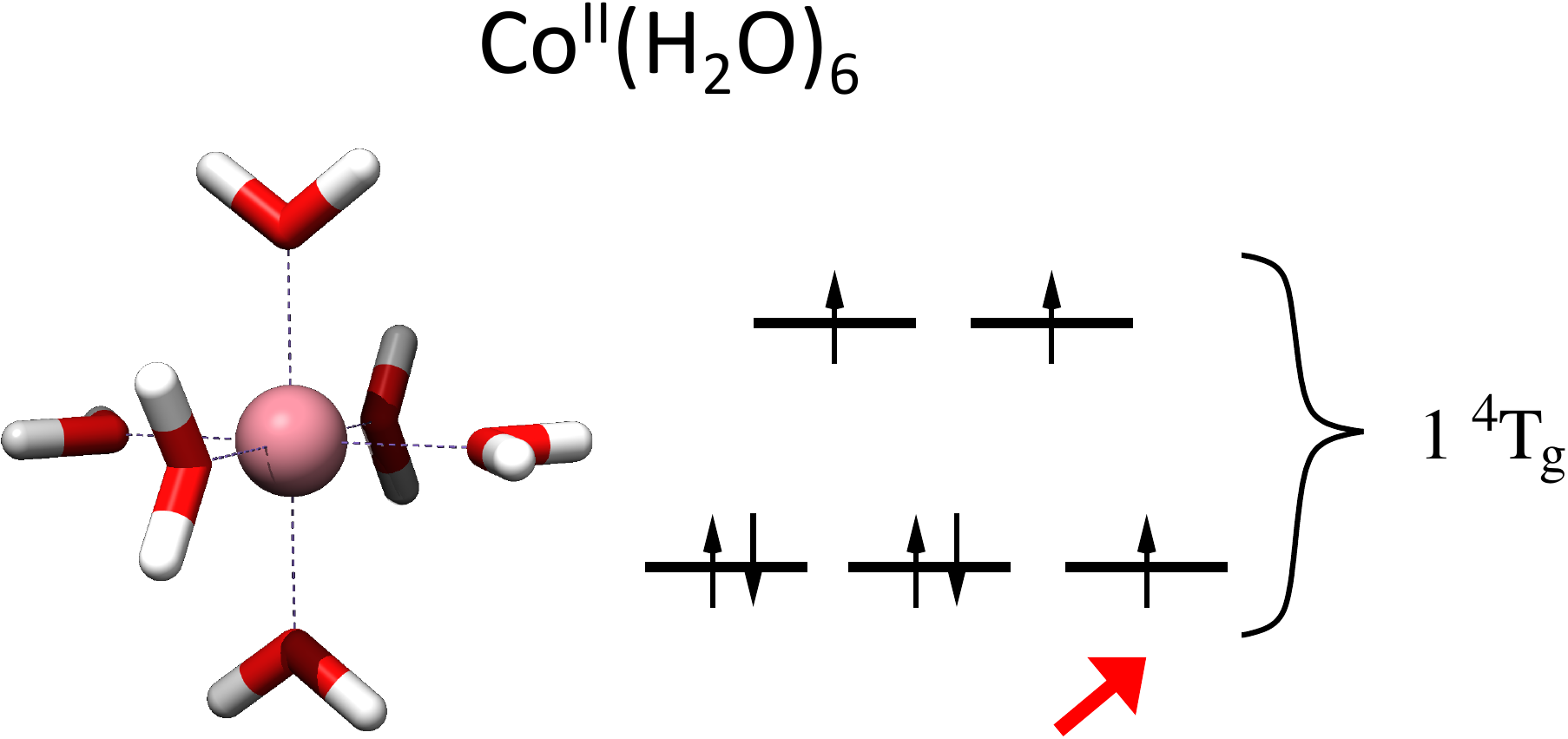}
   \caption{}
   \label{fig:5b}
	\end{subfigure}
  \begin{subfigure}{.48\textwidth}
   \centering
   \includegraphics[width=0.98\textwidth]{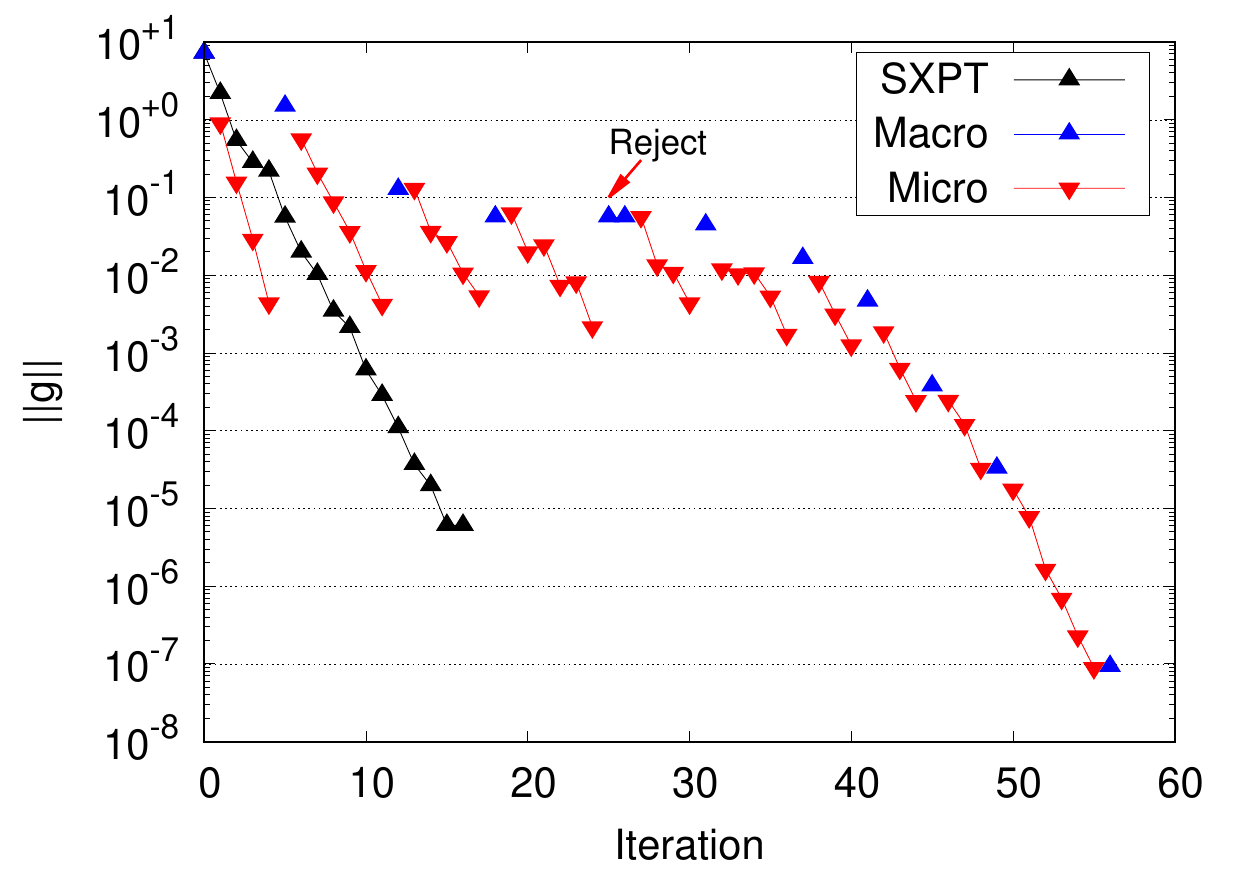}
   \caption{}
   \label{fig:5c}
  \end{subfigure}
  \begin{subfigure}{.48\textwidth}
   \centering
   \includegraphics[width=0.68\textwidth]{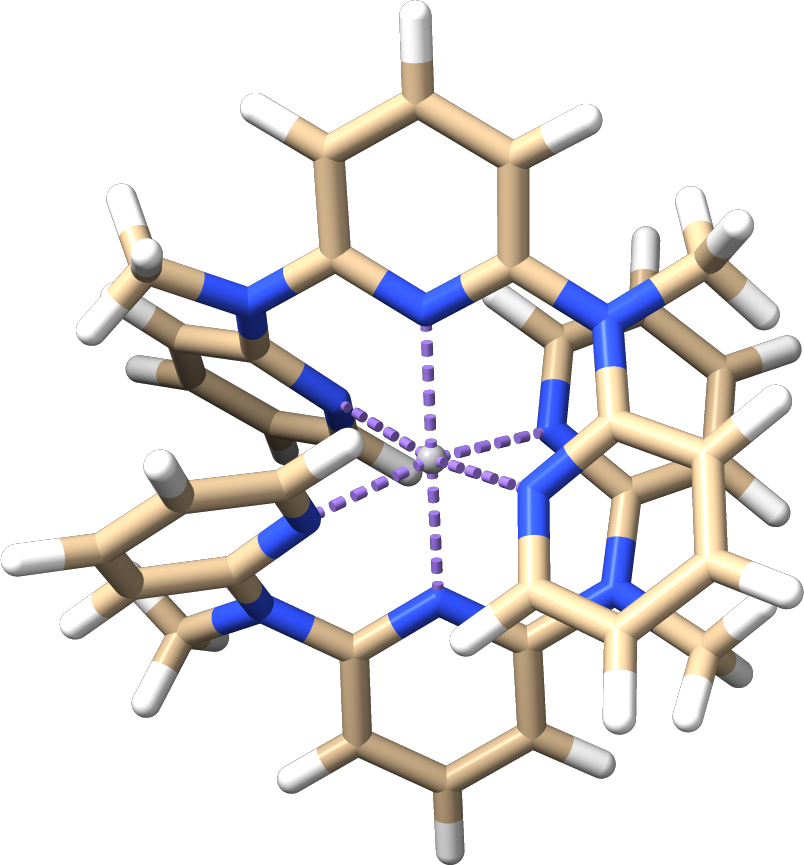}
   \caption{}
   \label{fig:5d}
	\end{subfigure}
 \end{subfigure}
\caption{
SXPT and TRAH convergence of SA-CASSCF/def2-TZVPP for the degenerate $1~{^4}T_g$ state (\subref{fig:5a})
 of [Co(II)(H\tief{2}O)\tief{6}]\hoch{2+} (\subref{fig:5b})
and the quasi-degenerate $1~{^3}T_g$ state (\subref{fig:5c})
 of a larger V(III) complex\cite{Dorn2020} (\subref{fig:5d}).
 Each calculation used a minimal 3d TM center active space together with the AVAS guess.
}
 \label{fig:5}
\end{figure}

%--------------------------------------------------------------------%
The SA-CASSCF approach is often used in practice to describe the
 ground state of degenerate or nearly degenerate open-shell TMCs.
To demonstrate the applicability of our new TRAH implementation
 for such calculations, the convergence is investigated for 
 the octahedral [Co(H\tief{2}O)\tief{6}]\hoch{2+} complex\cite{Neese2007} with an exactly degenerate
 ${^4}T_{g}$ ground state and a larger  V(III) complex\cite{Dorn2020} that
 has a distorted octahedral coordination and is in the quasi-degenerate 
  ${^3}T_{g}$ ground state (See Fig.\ \ref{fig:5}).
For both TMC, we employed the def2-TZVPP basis set and a minimal
 active space that contains the TM 3d orbitals and electrons.
The initial MOs are obtained from the model one-electron Hamiltonian and the 
 subsequent AVAS procedure in accordance with the desired active-space choice.
From Figs.\ \ref{fig:5a} and \ref{fig:5c}, it can be seen that
 TRAH also converges for such electronic structures without
 major difficulties or flaws, though for the V(III) complex one step
 has to be rejected because the trust region became too large.
Again, SXPT converges with significantly less iterations to the same
 solution as TRAH, which underlines that the AVAS guess is well
 suited for TMC CASSCF calculations and that the electronic structure
 is simple enough to be handled securely by the first-order SXPT.

%--------------------------------------------------------------------%
\subsection{Large molecules}\label{sec:largemol}
\begin{figure}
 \centering
%
% \begin{subfigure}{\linewidth}
%  \centering
%  \captionsetup{justification=centering}
%
  \begin{subfigure}[b]{.32\textwidth}
   \centering
   \includegraphics[width=0.70\textwidth]{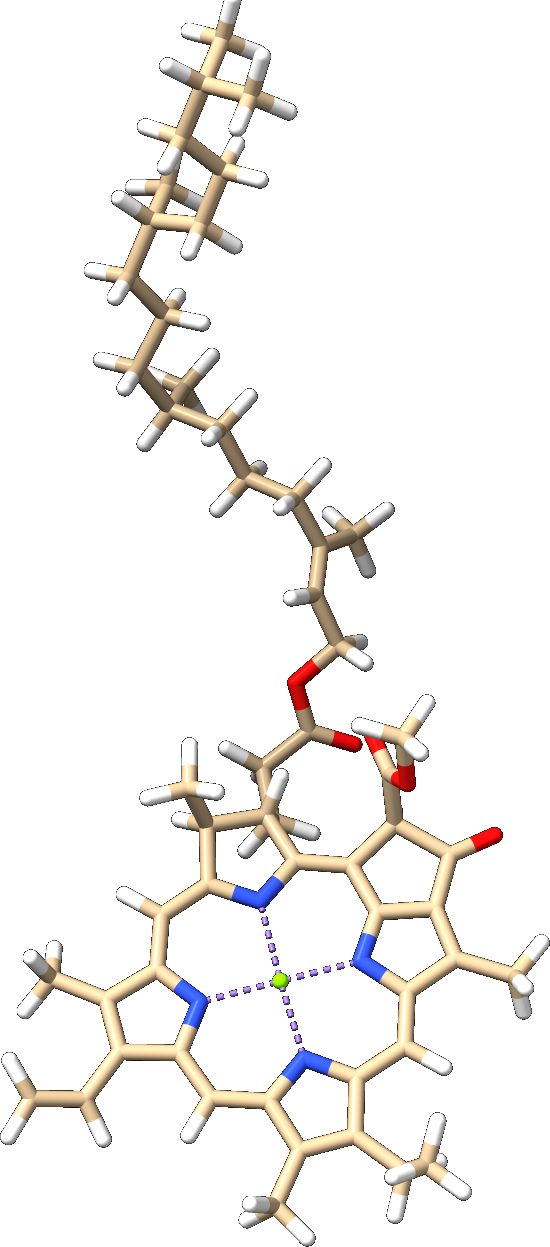}
   \caption{}
   \label{fig:6a}
  \end{subfigure}
  \begin{subfigure}[b]{.32\textwidth}
   \centering
   \includegraphics[width=0.98\textwidth]{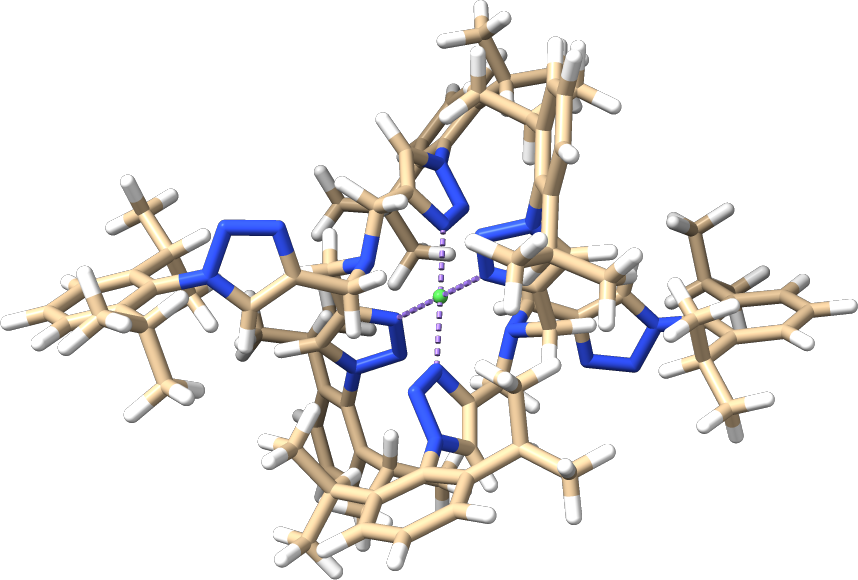}
   \caption{}
   \label{fig:6b}
  \end{subfigure}
  \begin{subfigure}[b]{.32\textwidth}
   \centering
   \includegraphics[width=0.98\textwidth]{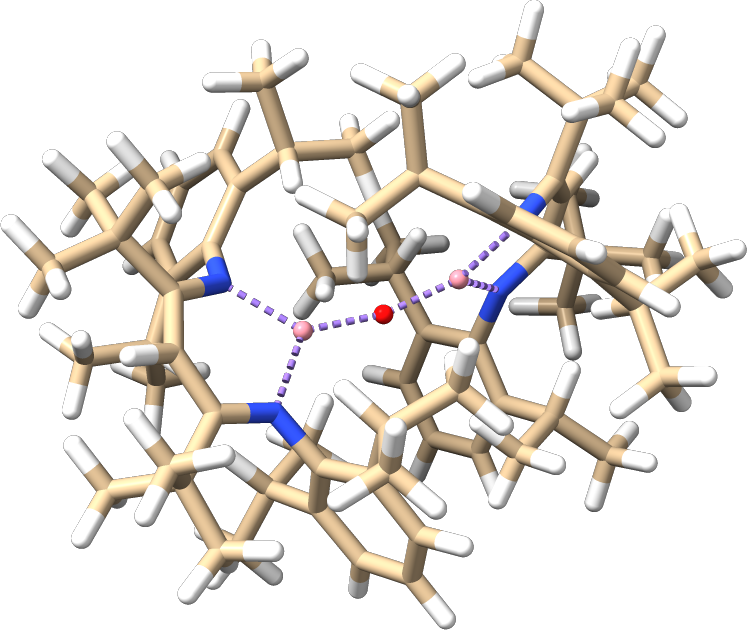}
   \caption{}
   \label{fig:6c}
  \end{subfigure}
%
 %\end{subfigure}
\caption{
 Structures of chlorophyll molecule (\subref{fig:6a}), Ni (\subref{fig:6b}) and Co (\subref{fig:6c}) complexes used in Sec.\ \ref{sec:largemol}. 
}
 \label{fig:6}
\end{figure}

%--------------------------------------------------------------------%
Finally, we would like to underline the applicability and usefulness
 of TRAH-CASSCF for large molecules.
For this purpose, the number of iterations and timings of SXPT and TRAH
 are given in Tab.\ \ref{tab:3}
 for three large molecules, which are shown in Fig.\ \ref{fig:6}
 and were employed previously by 
 other (approximate) second-order CASSCF solvers.\cite{Kreplin2020,Nottoli2021}
For the CAS(12,12)SCF calculation of the $1{^1}A$ state of chlorophyll,
 the cc-pVTZ basis set was employed as in Ref.\ \onlinecite{Nottoli2021}.
All chlorophyll calculations started from the UNO guess.
As shown in Fig.\ \ref{tab:3},
 both SXPT and TRAH implementations in ORCA converged in much less time ( 4 and 3 times faster, respectively)
 than the CD-NEO-CASSCF implementation.\cite{Nottoli2021}
It is likely that this factor would be even larger for serial calculations
 because we have used less MPI processes (16) than 
 the CD-NEO-CASSCF calculation OpenMP threads (28).
This runtime benefit for large systems is caused by the AO-driven
 implementation of Fock matrices that effectively scales linearly with the system size 
 for the costly exchange matrices when employing COSX.\cite{Neese2009,Izsak2011,Helmich-Paris2021b}
The CD-NEO-CASSCF implementation of Ref.\ \onlinecite{Nottoli2021}
 is MO based and scales as $\mathcal{O}(N^4)$ with the system size.
Such an MO-based CASSCF implementation is beneficial for small and medium-sized 
 molecules, as we have seen it in Sec.\ \ref{sec:res_arom},
 but it becomes a bottleneck for molecules with more than 100 atoms.
The same holds of course for our $\mathcal{O}(N^4)$ scaling RIJK implementation of Fock matrices.

The CASSCF total runtime either with SXPT or TRAH is only slightly larger
 than the time to prepare the UNO guess from a broken-symmetry calculation
 which took on the same machine with the same number of processes (16) altogether 2.00~h
 including two SCF calculations and one SCF stability analysis.
Hence, it remains debatable if the UNO guess from a BS-UHF calculation
 is the most reasonable start for large-molecule CASSCF calculations
 which require efficient implementations and computational work flows.

When looking at the runtime of the AO based SXPT and TRAH implementations,
 we notice again that the first-order SXPT implementation is faster than the 
 second-order TRAH implementation by factor of 1.4.
However, this runtime benefit does not reflect the much fewer number of SXPT (macro) iteration
 than the TRAH total number of iterations.
As mentioned in Sec.\ \ref{sec:impl_fock}, our TRAH implementation
 uses for the COSX matrices much smaller grids in the micro-iterations than in the macro-iterations.
This usually makes the costs of a single micro-iteration much smaller than the costs of 
 a macro-iteration provided that 
(i) the runtime is determined the Fock matrix builds, (ii) COSX is used, (iii) the molecule is sufficiently large,
 and (iv) the number of SA states is small.

Concerning the final CASSCF solution obtained with the SXPT and TRAH 
 calculations, we note that both calculations ended with 12 active orbitals
 that are purely of $\pi$ character and are all located in the chlorin chromophore.
Though TRAH found a solution with a lower energy, it is not clear
 how to rate these results eventually because both calculations gave a physically
 reasonable solution. 
Further CASSCF calculations using large active spaces that include all the $\pi$ orbitals and
 electrons in the active space would be desirable and worth to be investigated
 in more detail but, unfortunately, are currently not accessible, at least with
 a full second-order algorithm like TRAH.

% Ni(II) complex
The largest molecule (231 atoms) that is covered in this work is a Ni(II) complex\cite{Schweinfurth2013} (see Fig.\ \ref{fig:6}).
For the CASSCF calculations on the $1{^3}A$ state, a Ni(II) double d-shell active space, CAS(8,10),
 was chosen using the AVAS procedure.
To compare our results with those of the SOSX implementation\cite{Kreplin2020}, the def2-TZVPP
 basis set was employed.
From the results in Tab.\ \ref{tab:3}, it can be seen that both SXPT and TRAH converge to the same
 solution but the first-order SXPT implementation is 1.8 times faster than the full second-order TRAH method.
This indicates that the Ni(II) complex has a fairly simple electronic structure that can also
 be easily found by SXPT if a reasonable initial MO guess is employed (by AVAS) that is in accordance with the active space choice.
Furthermore, we notice that SOSX\cite{Kreplin2020} is 1.5 times faster than SXPT, which can be expected. 
SOSX employs the full orbital Hessian for all active-orbital rotations ($\kappa_{ti}$ and $\kappa_{at}$) 
 because the corresponding full four-center integrals with at least two active orbitals
 can be easily hold in memory and the NR micro-iterations are solved quickly.
In comparison to first-order algorithms, such as SXPT, 
 this implementation strategy provides more accurate wave function update vectors and, thus, leads to faster convergence.
The energy difference between SOSX and SXPT (and TRAH) is most probably related to
 the different integral approximations that were employed.
Though tight integral screening thresholds and grids were employed for our calculations,
 those differences should be considered and examined more closely when comparing with 
 experimental results or with more accurate wave function methods.

% Co-O-Co complex
Our final test example to study CASSCF convergence and runtime performance
 is a large spin-coupled dicobalt(II) oxo complex\cite{Roy2019}.
We employed a minimal active space with the 3d Co orbitals and initial MOs from 
 the model one-electron Hamiltonian processed by the AVAS procedure.
The def2-TZVPP basis set was employed again in order to compare with the SOSX results of Ref.\ \onlinecite{Kreplin2020}.
Though the SXPT calculation can converge the orbitals, it is not able 
 to converge the CAS-CI eigenvalue equations within 1024 iterations.
The TRAH calculation employs again localized active orbitals that turned out
 to be advantageous for spin-coupled systems (see Sec.\ \ref{sec:spin_coupl}).
Nevertheless, the number of TRAH micro-iterations is unusually large which
 results in a runtime of 43.9~h which is 6.5~slower than the SOSX
 calculation.
Still, TRAH-CASSCF securely finds a reasonable minimum solution with
 active orbitals located at the two Co(II) centers which SXPT could not provide.
As for the Ni(II) complex calculations, we notice again a small milli Hartree
 energy difference between the SOSX calculation\cite{Kreplin2020} and our
 TRAH-CASSCF calculation which is probably related to different integral approximations
 and screening thresholds.

%--------------------------------------------------------------------%
\section{Conclusions} \label{sec:concl}
%--------------------------------------------------------------------%

In this work, we introduced the step-restricted one-step full second-order TRAH-CASSCF
 converger for SS- and SA-CASSCF wave functions.

To avoid singularities in the configuration-configuration Hessian,
  an exponential parametrization of the CI coefficient updates is
  chosen that operates in a nonredundant orthogonal complement basis.
This parametrization is customary for SS-CASSCF response properties.
In this work, we have employed this parametrization for SS- and extended it for SA-CASSCF wave functions.
Within this formalism, MO and CI coefficients are updated in a similar 
 fashion using matrix exponentials of real anti-symmetric rotation parameters.
This is the cornerstone of a one-step algorithm and will be employed
 in future implementations of MR derivative-based properties.

We also completely avoid the costly intra-state rotations that
 occur for unequal-weight SA calculations by rotating
 the CI coefficients in a basis that diagonalizes the state-interaction Hamiltonian
 for those types of calculations.
 
Our implementation is integral-direct and atomic orbital-based,
 which facilitates calculations on large molecules that are still feasible for first-order solvers as well.
Our largest calculation, i.e. a CAS(8,10)-SCF/def2-TZVPP calculation of a Ni(II) complex with 231~atoms,
 finished within 44~h using 16~MPI processes.
In comparison with other state-of-the-art second-order implementations,
 our TRAH implementation seems to be more efficient than CD-NEO-CASSCF\cite{Nottoli2021} for large systems
 but slower than the approximate second-order SOSX method.\cite{Kreplin2020}
Also, we have compared the efficiency of TRAH-CASSCF with the first-order SXPT method\cite{Kollmar2019}.
In most cases, the first-order SXPT method is faster than TRAH simply because the total number
  of iterations is much smaller.
Nevertheless, we have seen in many cases that SXPT can have convergence issues
 even if the active space is reasonably chosen and reasonable initial MOs were provided.
In those calculations, TRAH-CASSCF always converged reliably, which is the main
 objective of a second-order converger.

In the course of this work, a few technical aspects became important to securely
 reach convergence with TRAH- as well as with SXPT-CASSCF:
(i) Canonical inactive and natural active orbitals generally improve the convergence rates
 of the micro-iterations.
However, for spin-coupled systems, localized active orbitals are preferable.
(ii) Large errors due to the RI approximation for active-orbital two-electron integrals
 can be easily avoided by a pivoted Cholesky decomposition of the Coulomb metric.
(iii) Primarily for TMC calculations, the S-fitted COSX can give severe numerical
 errors that can be drastically reduced by a variant that fully preserves the
 eight-fold permutational symmetry of the two-electron integrals at the
 expense of a minor additional computational overhead.

In future work, we will focus on reducing the total number of TRAH iterations
 by accounting for higher-order terms in the minimum energy Taylor expansion.
Other avenues to explore are CASSCF response properties for SA-CASSCF wave functions,
 such as excitation energies\cite{Helmich-Paris2019} and spin-Hamiltonian parameters for
 electron-paramagnetic resonance spectroscopy.

%--------------------------------------------------------------------%
\section{Acknowledgments}
%--------------------------------------------------------------------%

The author gratefully acknowledges financial support from Max Planck Society
 and the German Research Foundation (DFG) through grant no. HE 7427/4-1.
Moreover, the author cordially thanks Frank Neese and Kantharuban Sivalingam
 for fruitful scientific discussion.
The author is also indebted to Lisa Rosy and Shenfa Ye for providing
 coordinates of the dicobalt(II) oxo complex of Ref.\ \citenum{Roy2019}
 and Spencer L{\'e}ger for comments on the manuscript.

%\bibliography{references}
%aipnum4-2.bst 2019-01-14 (MD) hand-edited version of apsrev4-1.bst

%----------------------------------------------------------------------%
% fig 1
%----------------------------------------------------------------------%
\newpage

%----------------------------------------------------------------------%
% fig 2
%----------------------------------------------------------------------%
\newpage

%----------------------------------------------------------------------%
% fig 3a-d
%----------------------------------------------------------------------%
\newpage

%----------------------------------------------------------------------%
% fig 4
%----------------------------------------------------------------------%

%----------------------------------------------------------------------%
% fig 5
%----------------------------------------------------------------------%

%----------------------------------------------------------------------%
% fig 6
%----------------------------------------------------------------------%

%----------------------------------------------------------------------%
% tab 1
%----------------------------------------------------------------------%
\newpage
 \begin{table}[h]
 \caption{Intermediates of the atomic-orbital based gradient and sigma vector calculation}
 \label{tab:1}
 \begin{ruledtabular}
 \begin{tabular}{rcl}
 \multicolumn{3}{c}{Overlap and Reduced density matrices (RDM)} \\[0.5em]
 $\braket<0|0>$ & $=$ & $1$\\[0.4em]
 $D_{vw}$  & = & $\sum_i w_i \bra{0_i} \EOp_{vw} \ket{0_i}$ \\[0.4em]
 $d_{tu,vw}$ & = & $\sum_i w_i \bra{0_i} \E2Op_{tuvw} \ket{0_i}$ \\[0.4em]
 $\overline{\braket<0|S>}$ & $=$ & $0$ \\[0.4em]
 $\overline{D}_{vw}$  & = & $- \sum_j w_j \left( \bra{0_j} \EOp_{vw} \ket{S_j} + \bra{S_j} \EOp_{vw} \ket{0_j} \right)$ \\[0.4em]
 $\overline{d}_{tu,vw}$ & = & $-\sum_j w_j \left( \bra{0_j} \E2Op_{tuvw} \ket{S_j} + \bra{S_j} \E2Op_{tuvw} \ket{0_j} \right)$ \\[1.0em]
 \multicolumn{3}{c}{Orbital gradient} \\[0.5em]
 $F^X_{pq}$ & $=$ & $\sum_{\mu\nu} C_{\mu p} F^X_{\mu\nu} C_{\nu q}$ \\[0.4em]
 $F^X_{\mu\nu}$ & $=$ & $\delta_{I,X} h_{\mu\nu} + G_{\mu\nu}[\bfD^X]$ \\[0.4em]
 $G_{\mu\nu}[\bfD^X]$ & $=$ & $\sum_{\gk\gl} \left[ (\mu\nu|\gk\gl) - \frac{1}{2} (\mu\gl|\gk\nu) \right] D^X_{\gk\gl}$ \\[0.4em]
 $D^I_{\mu,\nu}$ & $=$ & $2 \, \sum_{k} C_{\mu k} C_{\nu k}$ \\[0.4em]
 $D^A_{\mu,\nu}$ & $=$ & $\sum_{vw} C_{\mu v} D_{vw} C_{\nu w}$ \\[0.4em]
 $Q_{pt}$  & = & $\sum_{uvw} (pu|vw) d_{tu,vw}$ \\[1.0em]
 \multicolumn{3}{c}{Configuration gradient} \\[0.5em]
 $E^c$ & = & $E_{\text{nuc}} + \frac{1}{2} \sum_{\mu\nu} ( h_{\mu\nu} + F^I_{\mu\nu} ) D^I_{\mu\nu}$ \\
 $h'_{tu}$ & = & $F^I_{tu}  - \frac{1}{2} \sum_v (tv|vu)$ \\[1.0em]
 \multicolumn{3}{c}{Orbital-orbital sigma} \\[0.5em]
 $\widetilde{F}^X_{pq}$ & = & $\sum_{\mu\nu} \left( \Lambda_{\mu p}  C_{\nu q} + C_{\mu p}  \Lambda_{\nu q} \right) F^X_{\mu\nu}$  \\[0.4em]
 & & $+ \sum_{\mu\nu} C_{\mu p}  C_{\nu q} \, G_{\mu\nu}[\widetilde{\bfD}^X]$ \\[0.4em]
 $\widetilde{D}^I_{\mu,\nu}$ & = & $2 \, \sum_{k} \left( \Lambda_{\mu k}  C_{\nu k} + C_{\mu k}  \Lambda_{\nu k} \right)$ \\[0.4em]
 $\widetilde{D}^A_{\mu,\nu}$ & = & $\sum_{vw} \left( \Lambda_{\mu v} D_{vw} C_{\nu w} + C_{\mu v} D_{vw} \Lambda_{\nu w} \right)$ \\[0.4em]
 $\widetilde{Q}_{pt}$ & = & $\sum_{uvw} \widetilde{(pu|vw)} d_{tu,vw}$ \\[1.0em]
 \multicolumn{3}{c}{Orbital-configuration sigma} \\[0.5em]
 $\overline{F}^A_{pq}$ & = & $\sum_{\mu\nu} C_{\mu p}  C_{\nu q} \, G_{\mu\nu}[\overline{\bfD}^A]$ \\[0.4em]
 $\overline{D}^A_{\mu,\nu}$ & = & $\sum_{vw} C_{\mu v} \overline{D}_{vw} C_{\nu w} $ \\[0.4em]
 $\overline{Q}_{pt}$  & = & $\sum_{uvw} (pu|vw) \overline{d}_{tu,vw}$ \\[1.0em]
 \multicolumn{3}{c}{Configuration-orbital sigma} \\[0.5em]
 $\tilde{E}^c$ & = & $\frac{1}{2} \sum_{\mu\nu} (\, (h_{\mu\nu} + F^I_{\mu\nu} ) \tilde{D}^I_{\mu\nu} + G_{\mu\nu}[\widetilde{\boldsymbol{D}}^I] D^I_{\mu\nu} \, )$ \\[0.4em]
 $\tilde{h}'_{tu}$ & = & $\tilde{F}^I_{tu}  - \frac{1}{2} \sum_v \widetilde{(tv|vu)}$
\end{tabular}
 \end{ruledtabular}
\end{table}

%----------------------------------------------------------------------%
% tab 2
%----------------------------------------------------------------------%
\newpage
 \begin{table}[h]
 \caption{
Timings of the SS CASSCF/cc-pVTZ calculations with CD-NEO\cite{Nottoli2021}, SXPT and TRAH using RIJK
 and the UNO guess.
Calculations ran on a single cluster node, 
 CFOUR calculations ran with 28 threads,\cite{Nottoli2021} 
 ORCA with 20 MPI processes on an Intel Haswell node (Intel{\textregistered} Xeon{\textregistered} CPU E5-2687W v3 @ 3.10~GHz) node.
}
 \label{tab:2}
 \begin{ruledtabular}
 \begin{tabular}{lrrrr}
Molecule       & CAS & \multicolumn{3}{c}{time ( min )} \\ 
               &     & CD-NEO & SXPT & TRAH \\
\hline\\
2Me2HSDiox     & ( 4, 4) &  0.76 &  0.93 &  1.35 \\
2Me4HSDiox     & ( 6, 6) &  0.61 &  0.58 &  0.79 \\
adrenaline     & ( 6, 6) &  1.04 &  1.47 &  1.96 \\
anthracene     & (14,14) &  8.27 & 19.35 & 16.36 \\
azulene        & (10,10) &  0.57 &  0.52 &  0.97 \\
biphenyl       & (12,12) &  0.93 &  1.63 &  2.25 \\
catechol       & ( 6, 6) &  0.11 &  0.27 &  0.42 \\
coumarin       & (12,12) &  4.55 &  4.51 &  6.96 \\
dopamine       & ( 6, 6) &  0.63 &  0.81 &  1.24 \\
fluorene       & (12,12) &  0.95 &  1.77 &  2.65 \\
indole         & ( 8, 8) &  0.31 &  0.44 &  0.67 \\
l-dopamine     & ( 6, 6) &  1.08 &  0.98 &  1.39 \\
naphthalene    & (10,10) &  0.34 &  0.55 &  1.07 \\
niacin         & ( 6, 6) &  0.14 &  0.42 &  0.66 \\
niacinamide    & ( 6, 6) &  0.17 &  0.48 &  0.76 \\
nicotine       & ( 6, 6) &  0.85 &  1.34 &  1.90 \\
noradrenaline  & ( 6, 6) &  0.74 &  1.00 &  1.48 \\
picolinic acid & ( 6, 6) &  0.16 &  0.34 &  0.67 \\
pyridine       & ( 6, 6) &  0.05 &  0.12 &  0.20 \\
pyridoxal      & ( 8, 8) &  0.88 &  1.07 &  1.42 \\
pyridoxamine   & ( 6, 6) &  1.21 &  1.22 &  1.66 \\
pyridoxine     & ( 6, 6) &  0.86 &  1.09 &  1.60 \\
resveratrol    & (14,14) & 11.43 & 12.84 & 11.74 \\
serotonin      & ( 8, 8) &  0.95 &  1.48 &  1.97 \\
tryptophan     & ( 8, 8) &  2.08 &  2.12 &  2.91
 \end{tabular}
 \end{ruledtabular}
\end{table}

%----------------------------------------------------------------------%
% tab 3
%----------------------------------------------------------------------%
\newpage
 \begin{table}[h]
 \caption{
Runtime performance and total energies of SS-CASSCF calculations 
 for three large molecules.
Calculations ran on a single cluster node (two AMD EPYC\hoch{\texttrademark} 7302 16-Core processors)
 with 16~MPI processes.
}
 \label{tab:3}
 \begin{ruledtabular}
 \begin{tabular}{lrrrr}
Algo & Macro It. & Reject & time (h) & energy (a.u.) \\ 
    & (Total It.) & & &  \\
 \hline \\[.4em]
\multicolumn{5}{c}{MgC\tief{55}N\tief{4}O\tief{5}H\tief{72}, CAS(12,12)} \\[.5em]
NEO\footnotemark[1]  & 12  &    & 12.25 & /            \\
SXPT &  33  &    & 2.49 & -2917.547 110 \\
TRAH & 12 (111) & 1 & 3.52 & -2917.556 419 \\[1.0em]
\multicolumn{5}{c}{[NiC\tief{90}N\tief{20}H\tief{120}]\hoch{2+}, CAS(8,10)} \\[.5em]
SOSX\footnotemark[2] &  17  & & 10.9  & -6074.966 405 \\
SXPT & 32  & & 15.49 & -6074.970 621 \\
TRAH & 20 (153) & 4 & 27.43 & -6074.970 621 \\[1.0em]
\multicolumn{5}{c}{Co\tief{2}ON\tief{4}C\tief{70}H\tief{106}, CAS(14,10)} \\[.5em]
SOSX\footnotemark[2] & 35 & & 6.8 & -5768.898 833 \\
SXPT & DNC & & / & / \\
TRAH & 27 (278) & 4 & 43.9 & -5768.901 217
 \end{tabular}
 \end{ruledtabular}
 \footnotetext[1]{Data taken from Ref.\ \onlinecite{Nottoli2021}}
 \footnotetext[2]{Data taken from Ref.\ \onlinecite{Kreplin2020}}
\end{table}

\end{document}